\documentclass[5p,times]{elsarticle}
\usepackage{lineno,hyperref}
\usepackage{latexsym,amsmath,amsfonts,amssymb,amsbsy,natbib,epsfig}
\usepackage{graphicx}
\usepackage{caption}
\usepackage{subcaption}
\usepackage{stfloats} 
\usepackage{listings}
\usepackage{ulem}

\newcommand{\C}{\uline}
\newcommand{\Li}{\dotuline}

\journal{}

\nolinenumbers

\begin{document}

\begin{frontmatter}

\title{Can I do it?}

\author{Joris Sijs}\corref{mycorrespondingauthor}
\cortext[mycorrespondingauthor]{Corresponding author}
\ead{j.sijs@tudelft.nl}
\author{Carlos Hernandez-Corbato}
\address{University of Technology, Mekelweg 26, Delft, The Netherlands}
\author{Willeke van Vught, Julio Oliveira}
\address{TNO, Oude Waalsdorperweg 63, Den Haag, The Netherlands}




\begin{abstract}
Knowledge about how well a robot can perform a specific task is currently present only in engineering reports which are inaccessible to the robot. Artificial Intelligence techniques, such as hypergraphs and automated reasoning, can provide such engineering knowledge online while enabling updates in the knowledge with new experiences. This requires a sound knowledge structure and maintenance routines for keeping this knowledge-base about the robot's capabilities truthful. A robot with such up-to-date information can reason about if and how well it can accomplish a task. This article introduces a knowledge representation that combines an ontology on system engineering, a deductive reasoning on the connections between system components, and an inductive reasoning on the performance of these components in the current system configuration. This representation is further used to derive the expected performance for the overall system based on a continuous evaluation of the actual performance per component. Our real-life implementation shows a robot that can answer questions on whether it can do a specific task with the desired performance.
\end{abstract}

\begin{keyword}
ontology, automated reasoning, performance and reconfiguration
\end{keyword}

\end{frontmatter}

\nolinenumbers

\section{Introduction}
The typical use of robotic systems is extending from closed environments, such as a workcell, to ``open world'' environments that are complex and unpredictable, such as factory plants and family residences. Also, expectations are that robotic systems will not anymore conduct a single task, see \cite{huizing}. Instead, they are prepared to suit a more general purpose in which they need to conduct a series of tasks, and where completely new task should not be excluded a-priori \cite{Russel}. As a consequence, one can expect that during its lifetime a robotic system is asked to perform new types of tasks and that the span of situations under which it is operating widens. Therefore, the system should ask itself: can I do this task?

\begin{figure}[!ht]
    \vspace{0cm}
        \centerline{\includegraphics[width=0.95\columnwidth]{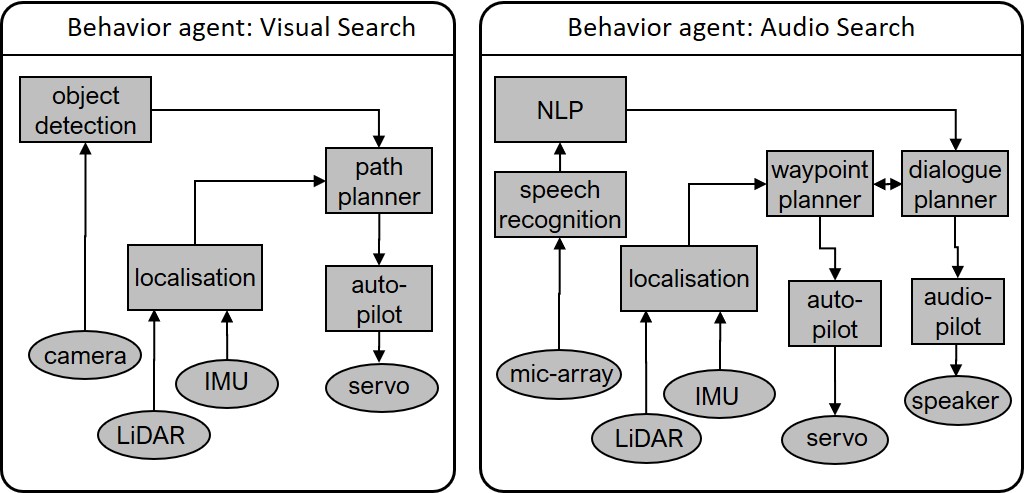}}
        \caption{Two alternative system configurations to search for persons.}
    \label{fig_search_visual_acoustic}
    \end{figure}

This ability, of a system that performs some kind of internal assessment, is often studied in the context of self-adaptation and self-governance of systems and, more recently, in model-based system engineering also. See, for example, the work in \cite{Pickery}, \cite{Heylighen} and \cite{chapman2018engineering}. Such studies assume that an initial design of hardware and software components is available, but it can be reconfigured by the system itself. That is, the system can further adjust settings of its software components, such as parameters and computational models, and adjust the system configuration that its components are in, i.e., their input-output connections. To illustrate the use of self-reconfiguration, let us assume that a fire brigade has a robotic system that can search for victims in a house and approach them for further assistance. Usually, such a system will invoke a single configuration running all the components under predefined settings by which it can conduct all of its tasks adequately in any conceived situation. This fit-to-all configuration is not necessarily optimal.  Given a specific situation and task, there are often other, less exhausting and more effective configurations. An example are the two possible configurations for a search task in Figure~\ref{fig_search_visual_acoustic}. One visual search conducted with a camera and visual object-detection that is effective in daylight condition when visbility is good; and one audio search with a microphone, speech recognition and Natural Language Processing that is more effective in dark and quite conditions when visibility is bad and noise levels are low. The figure illustrates the actual challenge: a system that is able to choose its configuration so that the execution of a planned task is most likely to succeed under the current conditions. 

In this article an approach is presented, called Self-X, that assesses all known configurations for completing a task and selects only those that match the operational conditions. Our approach introduces an updatable and online knowledge-base structuring configurations, tasks and operational conditions according to an ontology. The main benefit when using an ontology to structure the knowledge-base is that enables future interfaces with engineering languages as SysML and allows the system to perform further inference. Online routines assessing specific system components then interact with this knowledge-base so that the system may keep track of the quality, or accuracy, of any output currently produced by such components. This enables the system to make queries such as: 
\begin{itemize} 
    \item Which available configurations match the current conditions by which the task can be executed?
    \item Which available configurations satisfy the accuracy, or performance, that should be accomplished by the task?
\end{itemize}
With the results of these queries the system may select an effective configuration for a task, i.e., producing the desired result.

The main contributions of our knowledge-base approach are: (i) an ontology for characterizing system engineering knowledge still relevant when the system is operational, (ii) an implementation of this ontology in a hypergraph that may be updated with actual information about the performance of internal components and, (iii) inference methods for analyzing the system configuration (deductive and inductive reasoning).

The article is organized as follows. Some background and related work is presented in Section~\ref{Section_Related_Work_And_Background}. Sections~\ref{Section_component} and~\ref{Section_behavior} present the ontology in the stages of a component and a configuration with details on how to use it for modeling actual configurations. After that, in Section~\ref{Section_RealLife_Example}, a real-life example of a robotic system is presented in which the developed ontology is further integrated with online modules assessing operational conditions and system performances. Section~\ref{Section_Experimental_Results} then presents the experimental results, followed by Section~\ref{Section_Conclusions} concluding with the main points of our approach as well as future research still necesary.

\section{Background and Related work}
\label{Section_Related_Work_And_Background}

\subsection{Background on system modelling}
Engineering new systems, such as a robotic system, is an interplay of decomposing, developping and testing different (sub) designs of a system starting from a set of requirements. A description of each design may be formalized in a system's design model, or system model. In model based systems engineering, for example in \cite{estefan2007survey, dori2016model} but also in engineering in general, there are four important design categories: hard- and software design, physical design, logical design and functional design. Modelling languages as SysML \cite{friedenthal2014practical} are often used to formalize such designs, also in robotics, e.g., \cite{Bruyninckx_MBSE}, and support a system engineer to manage the \textit{vast} decision space of a system's complete internal structure \textit{before} it is used in operation (development time). Self-adaptation however, is authorized to make \textit{some} decisions about the internal structure \textit{when} the system is in operation (runtime). Therefore, Self-X has a different purpose compared to SysML, yet shares many of its ideas.

\begin{figure}[!ht]
    \vspace{-0.12cm}
        \centerline{\includegraphics[width=0.95\columnwidth]{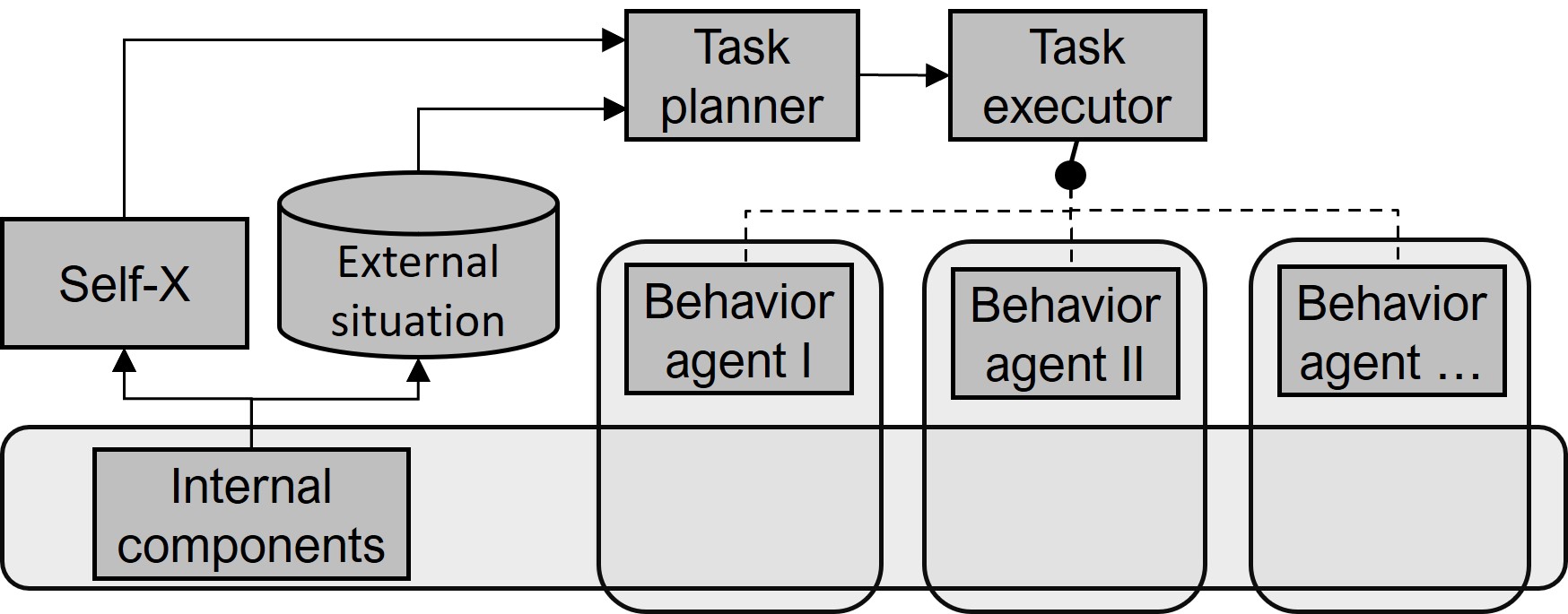}}
        \caption{Setup of an autonomous robot that is planning how to conduct a task as a series of different behaviors.}
    \label{fig_general_setup}
    \end{figure}

Most decisions in self-adaptation are related to the functional design of a system, which presents a functional model of the robotic system so that it will display certain behavior in order to complete the task it has recieved \cite{Lind94, Carlos_Tomasys_2018}. These functional models are often used in development time to provide guarantees that system requirements are met, for example in what the system is able to perceive from its noisy sensor readings and uncertain world models \cite{WeynsArxiv}. Some research started to focus on formal (functional) models providing similar guarantees for self-adaptive systems in runtime, see \cite{LemosGiese, IftikharWeyns, ZhangCheng}, where models were developed for self-organizing software agents operating in a virtual environment rather than autonomous robots operating in the real physical world. Still, their relevancy to Self-X is that such models indicate what structure of software agents provide which function of the system, while autonomous robots need similar models to assess what configuration of components provides the most effective behavior for completing its task under the current environmental conditions. Furthermore, a unique aspect of an autonmous robot is that they have a planning capability, for example as specified in the MAPE-K architecture found in \cite{MAPE_K} and the 4D-RCS reference architecture described in \cite{Schlenoff_4DRCS} often used to start the functional design of a robot. A high-level, task or mission planner will determine which of the available behaviors should be performed (in series or in parallel) in order to complete a task, while each behavior is executed by an particular agent. Here, a behavior agent should be perceived as the orchastrator of a (partial) system configuration of internal components that ensures the system will conduct such behavior, as was also illustrated in figure~\ref{fig_search_visual_acoustic}. Then, figure~\ref{fig_general_setup} depicts a typical setup of a robot in which a task (or mission) planner will schedule the behavior agents based on its perception of the external world. Yet, a key aspect of an \textit{autonomous} robot is that planning of a task should also be based on the robot's awareness about the internal workings of the system, i.e., the expected performance of the system configuration that is executing its current behavior as provided by Self-X. Functional models characterizing such performance are the cornerstone of self assessment. The Self-X approach described in this article offers a grounding of functional modeling with respect to the behaviors, tasks and environmental conditions of a robot, while the modelling is formalized as an ontology that conceptualizes the robot's engineering knowledge.

\subsection{Background on ontology}
The importance of an ontology in robotics is clearly visible with the standardization initiative on robotics and automation (CORA) committed by IEEE in \cite{CORA_2013} and  \cite{Neto_CORA_2019}. The benefit of an ontology is that an explicit structure of concepts and relations in the real world may be used by the robot for further inference to gain a deeper understanding of the robot's situation.

Recent studies extended the ontology of CORA to structure how a robot may execute a task. The extension starts with the task ontology found in \cite{CORA_TO_2017} introducing concepts \C{Robot}, \C{Task}, \C{Capability}, \C{Action} and \C{Behavior}. Each \C{Capability} describes an ordering of more detailed tasks, which again can be realized by (other) capabilities until the lowest level of a task is reached: the level of \C{Action}. An action is executed by the robot with one of its behaviors, where multiple behaviors may be possible to execute an action. The difference between \C{Behavior} and \C{Capability} is, similar to \C{Task} and \C{Action}, in the decomposition.

Other studies introduce an ontology to structure the actual environment in which the robot is operating, for example in~\cite{Waterloo_Part1, Waterloo_Part2} for intelligent vehicles, or in~\cite{Thai_Grun_2020} on how a robot could percieve and interact with the physical world. This work on ontologies for environmental modeling is often limited to one specific use case and in specific situations, i.e., a closed world. A less specific ontology related to an indoor setting with open world assumptions is the KnowRob system, first introduced in~\cite{KnowRob2010} and later extended to an updated version in~\cite{KnowRob2018}. KnowRob offers a heterogenous solution to semantics in the cognitive part of a robot operating in the household domain. Heterogenous in the sense that information of the environment from sensor observations is further stored in the knowledge base of the robot, so that this information is combined with contextual knowledge. Still, structuring the impact of external conditions to the internal configuration of a robotic system and updating such a structure, or ontology implementation, with the latest sensor readings has recieved little to no attention at all.

\subsection{Background on hypergraphs for designing an ontology}
\label{background_hypergraphs}
\begin{figure}[!ht]
     \centering
     \begin{subfigure}[b]{0.45\columnwidth}
         \centering
         \includegraphics[width=0.9\textwidth]{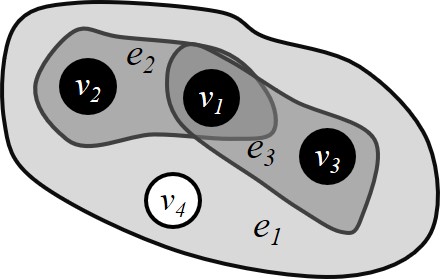}
         \caption{original representation}
         \label{fig_hypergraph_true}
     \end{subfigure}
     \begin{subfigure}[b]{0.5\columnwidth}
         \centering
         \includegraphics[width=0.9\textwidth]{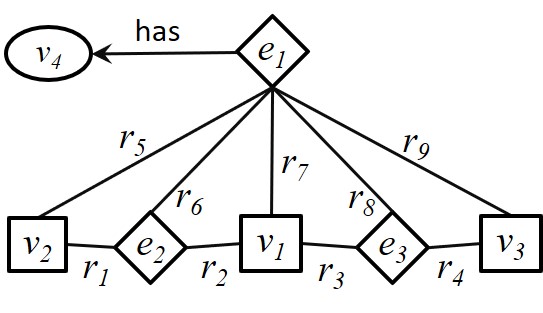}
         \caption{plain graph representation}
         \label{fig_hypergraph_notation}
     \end{subfigure}
        \caption{An illustrative example of a hypergraph, in its original representation (a), and in a plain graph representation (b).}
        \label{fig_hypergraph}
  \end{figure}

Robotic systems operate in an open, complex environment. Designing a knowledge structure, such as an ontology, that is able to capture such complexity as a plain graph, i.e., with edges linking two vertices, is often accompanied by a complex administration to maintain the proper structure \cite{ScioniBruyninckx2016}. This limitation can be removed when the structure of an ontology assumes a hypergraph design, for example the one in Figure~\ref{fig_hypergraph_true}. In a hypergraph an edge as $e_1$ can join multiple vertices as $v_1$, $v_2$, $v_3$ and $v_4$, while other edges between these vertices, such as $e_2$ and $e_3$, can also be part of the ``overarching'' edge $e_1$. For clarity, a hypergraph can be drawn as a plain graph (Figure~\ref{fig_hypergraph_notation}) by treating the hyper-edge as another type of vertices in the plain graph and add an argument $r_i$ to each of the plain edges.

This article adopts the plain graph representation of a hypergraph when developping its ontology. The ontology is derived from three (meta) concepts and two links, where a concept is either a hyper-vertices or hyper-edge and a link is a plain edge:
  \begin{itemize} 
	\item \C{Entity}: A subclass of \C{Concept} that can consist independent from any other concept ($v_1$, $v_2$, $v_3$ in Figure~\ref{fig_hypergraph});
	\item \C{Relation}: A subclass of \C{Concept} on how multiple entities and/or relations are joined ($e_1$, $e_2$, $e_3$ in Figure~\ref{fig_hypergraph});
	\item \C{Attribute}: A subclass of \C{Concept} \textit{with a value} modelling a property of an entity or relation ($v_4$ in Figure~\ref{fig_hypergraph});
	\item \Li{Role}: A subclass of \Li{Link} expressing the role that the entity or relation plays in a relation ($r_1$ until $r_9$ in Figure~\ref{fig_hypergraph});
	\item \Li{has}: A special subclass of \Li{Link} indicating that an attribute is owned by some entity, relation, or attribute. 
  \end{itemize}
With the concepts \C{Entity}, \C{Relation}, \C{Attribute}, and the links \Li{has}, \Li{Role}, the ontology specifies possible links between two concepts with the notation $\langle Concept, Concept \rangle: Link$. Links between concepts are inherited by subclasses. When the link is a \Li{Role} the ontology autogenerates both pointing options, while if the link is a \Li{has} it is unidirectional and pointing to an attribute:

$\langle Entity, Entity \rangle \! : \! Role$ \hspace{0.9cm} ; \hspace{0.2cm} $\langle Entity, Attribute \rangle \! : \! has$ \hspace{0.25cm} ;

$\langle Relation, Relation \rangle \! : \! Role$ \hspace{0.27cm} ; \hspace{0.2cm} $\langle Relation, Attribute \rangle \! : \! has$\hspace{0.1cm};

$\langle Attribute, Attribute \rangle \! : \! Role$ \hspace{0.1cm} ; \hspace{0.15cm} $\langle Attribute, Attribute \rangle \! : \! has$ \hspace{0.01cm};

$\langle Entity, Relation \rangle \! : \! Role$ \hspace{0.5cm} $\Leftrightarrow$ \hspace{0.1cm}  $\langle Relation, Entity \rangle \! : \! Role$\hspace{0.35cm};

$\langle Entity, Attribute \rangle \!: \! Role$ \hspace{0.42cm} $\Leftrightarrow$ \hspace{0.1cm} $\langle Attribute, Entity \rangle \! : \! Role$\hspace{0.3cm};

$\langle Relation, Attribute \rangle \! : \! Role$ \hspace{0.08cm} $\Leftrightarrow$ \hspace{0.05cm} $\langle Attribute, Relation \rangle \! : \! Role$.\\

In order to reason with hypergraphs standard notation from 1$^{st}$ order or description logic could be used, yet standard reasoning in hypergraphs would already require large and complex logic equations. Therefore, a new notation is introduced for creating 2$^{nd}$ order logic in hypergraphs, where instances of \C{Concept} are represented as a set allowing to use set theory. In set theory $\bar{A}$ denotes the compliment of a set $A$ (also known as the ``not'' in logic $\lnot A$), while $\cup$, $\cap$, $\setminus$ and $\Delta$ denote the union, intersection, set minus and symmetric difference of any two sets.

To represent concepts as a set, let $c \in$ Concept define an individual variable $c$ as being any instance of \C{Concept} that is present in the knowledge base. Also, let $C = \{\text{Concept}\}$ denote the definition $C := \cup \{c\}, \forall c \in$ Concept, i.e., $C$ is a set variable that collects all instances of \C{Concept} present in the knowledge base. Further, note that links are, in principle, not a set but as they point concepts a link can be defined by a set. Therefore, let $c.hasAttribute \subset \{ \text{Attribute} \}$ denote the set $c.hasAttribute$ that collects all instances of \C{Attribute} that are owned by a particular instance $c \in$ Concept. And let $c.Role \subset \{ \text{Concept} \}$ denote the set $c.Role$ that collects all instances of \C{Concept} that are linked to a particular instance $c \in$ Concept via \Li{Role}.

As an example, suppose that \C{Image} is a subclass of \C{Entity}, \C{Process} of \C{Relation}, \C{Name} of \C{Attribute} and that both \Li{Input} and \Li{Output} are a subclass of \Li{Role}. Then, $i \in$ Image denotes that $i$ is any instance of the type \C{Image}. Similarly, one could have $p \in$ Process and $n \in$ Name. Additional, intersting examples are:
  \begin{itemize} 
	\item Specifying $\langle Image, Name \rangle : has$ gives that $i.hasName$, where $i \in$ Image, is a set of all instances of type \C{Name} that are owned by the known images $i$. Further, $I = \{i \in \text{Image} | \, \text{``Picture''} \in i.hasName\}$ is a set of all instances of \C{Image} having the name Picture. And when \C{Image} also owns attributes such as \C{Height} and \C{Width}, then

$I = \{i \in \text{Image} | \, 600 \in i.hasHeight, 400 \in i.hasWidth\}$

is a set of all instances of \C{Image} that have a height of 600 and a width of 400 (pixels). Note that $I \! \subset \! \{ \text{Image} \}$.
	\item Specifying $\langle Process, Image \rangle \! : \! Output$, and automatically also $\langle Image, Process \rangle \! : \! Output$, gives two options. A first one $i.output$, where $i \in$ Image, representing a set of all the instances of type \C{Process} in which the known image $i$ is an output. And a second $p.Output$, where $p \in$ Process, representing a set of all the instances of type \C{Image} that play a role of output in the known process $p$.
  \end{itemize}

\subsection{Related work on self-assessment}
Selecting the correct system configuration is already quite an engineering challenge for one task, while autonomous robots should execute a number of different tasks under the changing operational conditions of the real (open) world. The most appropriate configuration of system components will not only be different from one task to another, but also from one operational condition to another. For that reason, capabilities as self-assess and self-manage have been studied, yet under different terms: self-awareness in \cite{Lewis_2019}, self-confidence in \cite{FamSec, Sweet}, self-adaptive in \cite{Carlos_Tomasys_2018} and meta-control in \cite{Carlos_MetaControl2020}. This section presents how the results from this literature is advantageous, or not, for the challenge addressed here: a system that is knowledgeable for online assessing the performance of its configuration.

A framework on self-awareness that is introduced in \cite{Lewis_2015}, and further studied in \cite{Lewis_2019}, is a set of layered control loops suiting self-organizing systems. The scientific roots of this framework are in psychology, due to which the framework is a mapping of the five Neisser’s levels on human self-awareness: stimulus awareness, interaction awareness, time awareness, goal awareness and meta-self-awareness. Though the work is interesting to gain new ideas, it is still very conceptual and many aspects of the framework are not (yet) practically feasible.

Methods related to self-confidence focus on the plan that was created by the autonomous robot (or agent). In \cite{FamSec} this self-confidence is defined as: "An agent’s perceived ability to achieve assigned goals within a defined region of autonomous behavior ...". A similar scope is taken in \cite{Sweet}, though, therein more emphasis is put on the validation of the data used for planning. Note that one important aspect of these methods is that they are centered around the planning capability of a system. In that sense it is too limited for our purpose as that also includes assessing system components that do not produce a plan.

Among methods on self-adaptive keen approaches have been developed using the Ontology for Autonomous Systems (see \cite{Bermejo_Oasys_2010}). Most notably is the Teleological and Ontological Model for Autonomous Systems (TOMASys) in \cite{Carlos_Tomasys_2018}. This ontology provides the concepts for modeling the functional knowledge of robotic systems and was applied to a meta-controller in \cite{Carlos_MetaControl2020} for managing the configuration of an underwater vehicle. TOMASys allows for automated, symbolic reasoning on whether the configuration could execute the task as planned. At the current stage, TOMASys has some drawbacks preventing it for being completely suitable to our challenge. Firstly, the ontology solely describes the internal functioning of a system to detect failures and cannot relate those to environmental conditions. Secondly, TOMASys can only perform logic checks on (numerical) quality attributes per component to set their (Boolean) availability. Nonetheless, TOMASys gives excellent inspiration.

Another interesting approach is RobustSENSE, see \cite{Tas_Robustsense_2017}, proposed for automated vehicles. It contains a performance-assessment-unit quanitfying the reliability of each component in its configuration by combining the environmental conditions with meta-information in a Bayesian network. The RobustSENSE approach is very interesting, for example the ideas of a Bayesian network, and they are in line with some of our challenges. However, there are also some drawbacks, mostly related to the fact that no general knowledge structure is used. As such, there is no higher level of understanding as to why or how the configuration would produce a correct execution of the task. More precisely, both the configuration of the components, as well as the task that the vehicle is to achieve, are fixed in RobustSENSE and determined during the design of the system. 

Our conclusion is that some methods for assessing and/or managing the performance of a system are still conceptual and do not mention practical implementations, such as \cite{Lewis_2019, FamSec, Sweet}, while others stick to symbolic reasoning \cite{Carlos_Tomasys_2018, Carlos_MetaControl2020} and do not exploit any actual, numerical readings from within the system. Nor do they address how the environment is effecting this performance. This last aspect is covered by RobustSENSE in \cite{Tas_Robustsense_2017}, though that approach lacks the concept of a general knowledge structure, due to which it is based on expert implementations making ad-hoc choices when designing the system.

\subsection{Approach on Self-X}
Our approach on Self-X starts with the modeling engineering knowledge in an ontology and implement that in an online knowledge base. Individual components of the robotic system are modeled with their requirements for operating properly and, in addition, for each output that a component produces quality metrics further specify the accuracy of this output. A real-life example at the end of the article illustrates how these metrics are updated from actual assessment modules. The next step in Self-X is to extend the ontology with a configuration of components and relate these configurations to the behaviors that the robot may need to conduct. Relating behaviors with configurations would then support the robot to learn the probility that a behavior is completed succesfully when given the quality metrics per component in that configuration (depending on the operational conditions). The robotic system can then select another system configuration, i.e., behavior, to respond upon loss of performance.

\section{Modelling of individual components}
\label{Section_component}Our design of the ontology starts, in Section~\ref{Section_component_ontology}, with the design pattern of individual components and the creations that they produce. Then, Section~\ref{Section_component_design_patterns} presents some illustrative examples on how to instantiate components, such as a camera and a detection algorithm. The next Section~\ref{Section_behavior} will further continue on these results and describe how they are used to model configurations that define a robot behavior.   

\subsection{Design pattern for a component}
\label{Section_component_ontology} 
\begin{figure}[!ht]
    \centering
        \centerline{\includegraphics[width=0.9\columnwidth]{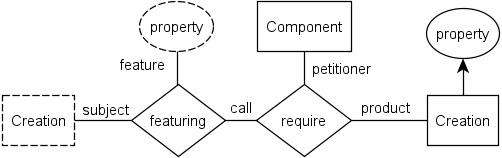}}
        \caption{The highest level concepts in the model for a Component producing a Creation while requiring other Creations to call for. In case a box has a dashed line it means that the concept is already present in the model and thus that, in the actual ontology and implementation, its edges should be connected to the original concepts having the solid line. However, when illustrating the ontology for a system engineering the above view is more clear compared to an illustration of the true ontology without such duplicates, as it illustrates a Creation at the 'input' and a Creation at the 'output' of the Component which is more in line with the domain of system engineering. Also note that a has-link is depicted as a closed arrow.}
    \label{fig_main_model}
    \end{figure}

The ontology starts with the most general characterization of any single (or set of) component(s) in the system's configuration. Hereto, as in model-based system engineering~\cite{Waymore}, a component is characterized as the concept \C{Component}, which is a type of \C{Entity}, and it is the most detailed level of engineering knowledge that is needed by the system. In general, a component requires one or more calls so that it can produce a \C{Creation} with particular properties, see also Figure~\ref{fig_main_model}, where \C{Creation} is a type of \C{Entity}. Each of these calls will again refer to a creation, but then a creation that features (not yet necessarily has) particular properties. In other words, our main model starts with a component that requires calls to one or more creations each featuring particular properties to produce another creation that has properties. Note that \Li{Require} and \Li{Featuring} are a type of (ternary) \Li{Relation}. Also, note that the role of \Li{Product} is played by the \C{Creation} that \Li{has} a set of properties. And the role of \C{Call} is played by the \Li{Featuring} relation which connects a \C{Creation} to some set of properties. The reason for this modelling decision is because a component might be open to call for creations within a range of properties, implying that, at this point, it should not already be limited to one specific creation with a specific property. Further explanations are provided in Section~\ref{Section_behavior}.

  \begin{figure*}[t]
     \centering
     \begin{subfigure}[b]{0.55\textwidth}
         \centering
         \includegraphics[width=1.0\textwidth]{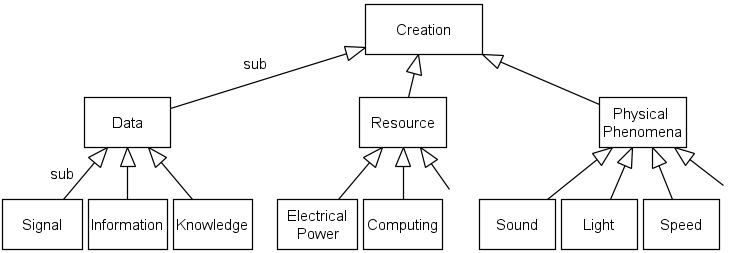}
         \caption{Hierarchy of the types of Creation}
         \label{fig_product}
     \end{subfigure}
     \begin{subfigure}[b]{0.4\textwidth}
         \centering
         \includegraphics[width=1.0\textwidth]{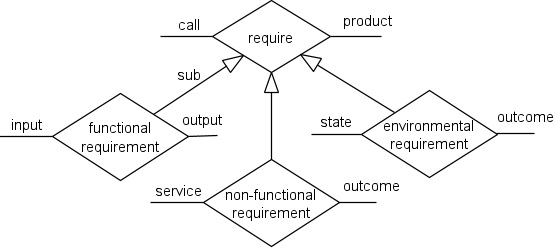}
         \caption{Hierarchy of the types of Require and its Roles}
         \label{fig_relying}
     \end{subfigure}
     \begin{subfigure}[b]{0.4\textwidth}
         \centering
         \includegraphics[width=1.0\textwidth]{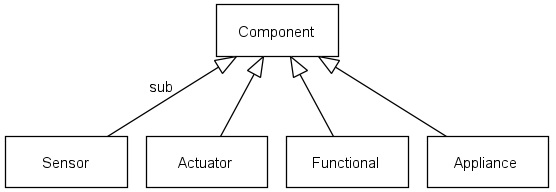}
         \caption{Hierarchy of the types of Component}
         \label{fig_unit}
     \end{subfigure}
     \begin{subfigure}[b]{0.5\textwidth}
         \centering
         \includegraphics[width=1.0\textwidth]{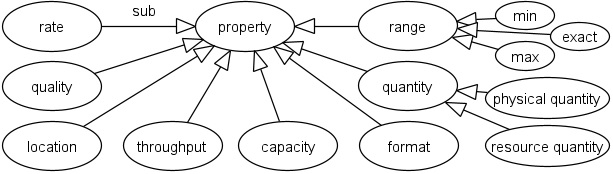}
         \caption{Hierarchy of the types of Property}
         \label{fig_properties}
     \end{subfigure}
        \caption{The hierarchy of sub-classes for Creation, Component, Require and Property. Every open-ended arrow is a sub-relation, while a dashed arrow implies that additional types can be included. Entities are illustrated as a rectangular box, relations as a diamond and attributes as an ellipse.}
        \label{fig_hierarchy_of_concepts}
  \end{figure*}
The next step is to further refine the concepts in Figure~\ref{fig_main_model} into sub-classes. These refinements should reflect our usage of the ontology, which is to understand a configuration and to adjust it if necessary. On that note, recall that adjustments typically act on software components of which there are two types: a functional component that computes or maps an analog or digital input to an output, and a driver component that interfaces with physical sensors and actuators. Robots also have components as batteries, communication systems and computing units, i.e., basic appliances that offer a resource to support other hardware and software components on the robot.  Therefore, the concept \C{Component} is refined as \C{Sensor}, \C{Actuator}, \C{Functional} and \C{Appliance}, see Figure~\ref{fig_unit}. A component produces a \C{Creation}, which can be some type of \C{Data} in the case of  a sensor or a functional (\C{Data} as \C{Signal}, \C{Information}, \C{Knowledge}); it can be some type of \C{Resource} in the case of an appliance (\C{Resource} as \C{Electrical Power}, \C{Computation}, \C{Communication}, etc.); or it can be a \C{Physical Phenomena} in the case of an actuator (\C{Light}, \C{Sound}, \C{Speed}, etc.). Note, that physical phenomena as light might also be present in the robot's environment and does not necessarily have to be produced by an actuator on the robot. These sub-classes of \C{Creation} are depicted in Figure~\ref{fig_product}. 

A component further requires other creations to call for. These creations may be the result of other components, such as data or a resource. In system engineering such requirements are known as (non-) functional requirements. In addition, components also depend on the conditions, or state, of the environment in which the system is operating, e.g., levels of  light, sound and visibility. Therefore, to model this, the \C{Require} of a component is refined as a \C{Functional Requirement} (FR), a \C{Non-Functional Requirement} (NFR), and an \C{Environemental Requirement} (ER) as is illustrated in Figure~\ref{fig_relying}. Note that, the role of \C{Petitioner} in this \C{Require} relation is not refined but the two roles of \C{Call} and \C{Product} are: \C{Call} is either \C{Input}, \C{Service} or \C{State}, while \C{Product} is refined into \C{Output} and \C{Outcome}. The difference between output and outcome is that an output is an actual thing produced by the component, such as data, while an outcome refers to the effect of how an internal resource or an external physical quantity will influence this output of the component.

  \begin{figure*}[b]
     \centering
     \includegraphics[width=0.8\textwidth]{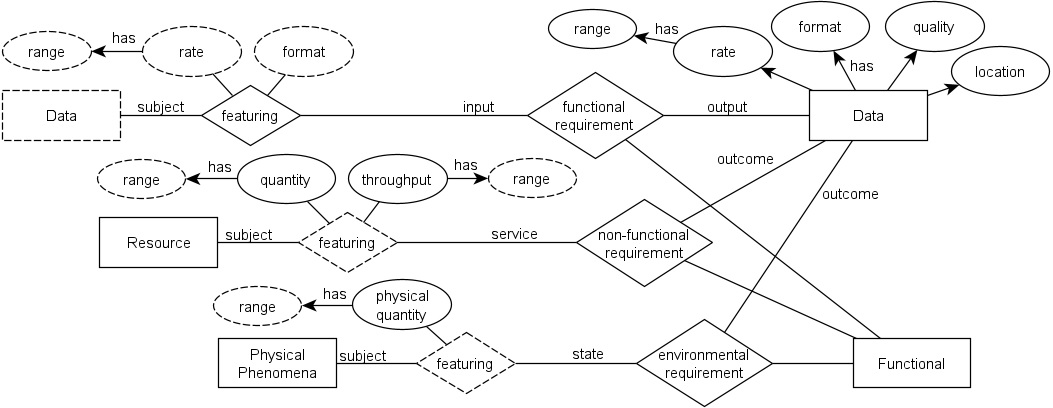}
     \caption{Ontology model of a Functional. Here, as previous illustration, a dashed line of a box indicates that the concept in that box is already present in another part of the model and that the dashad box is merely an illustrative copy of the original concept.}
     \label{fig_functional_unit_1}
  \end{figure*} 
   
Before continuing with the refinement of a property, let us illustrate how this first set of refinements on a component, its creation and what it requires, is used to characterize the difference between a sensor, actuator, functional and appliance.
   \begin{itemize} 
	\item \C{Sensor}: Turns a \C{Physical Phenomena}, that is a \Li{State} in the external world, into \C{Data} as an \Li{Outcome} (ER);
	\item \C{Actuator}: Turns \C{Data} at its \Li{Input} so as to \Li{Output} a \C{Physical Phenomena} in the robot's external world (FR);
	\item \C{Functional}: Takes \C{Data} as \Li{Input} and uses it to determine new \C{Data} at its \Li{Output} (FR);
	\item \C{Appliance}: Produces a \C{Resource} that may be used by another \C{Component} as a \Li{Service} (NFR).
  \end{itemize}
The last concept to be refined is \C{Property}, see Figure~\ref{fig_properties}, which on itself is a subclass of \C{Attribute}. To refine these properties one needs to understand how a system engineer would apply a FR, a NFR and the ERs. A FR describes what properties must be present on the input, so that the component can guarantee the properties of it's output. In case the input (or output) of a FR refers to data, then typical properties used in by system engineers are \C{Quality} (how accurate it is data), \C{Rate} (amount of data per unit of time), \C{Format} (the formatting type in which data is defined) and \C{Location} (where the data is found in the system). In case the output of a FR is a physical phenomena, then a typical property is the \C{Range}, e.g., \C{Min} and \C{Max}, between which the component is able to set the actual value of that phenomena in the real, physical world. Thereto, the phenomena is characterized with a \C{Physical Quantity} that is valued according to a physical unit. A NFR describes what properties are used to characterize a resource, such as \C{Capacity} (total amount of the resource that is available), \C{Throughput} (the amount of resource that can or should be provided per unit of time) and \C{Quantity} (a physical or resource quantity valued with a unit). Finally, an ER describes what property should be satisfied by physical phenomena in the robot's operational environment, which is typically some \C{Range} defined for a specific \C{Physical Quantity} that is valued in a physical unit. The concepts \C{Min} and \C{Max} and \C{Exact}, specifying a numerical level for an amount, are defined as a type of \C{Range} so that later on logic inference may take place such as: more than \C{Min}; equal to \C{Exact}; etc.

These refinements on creations, requirements and properties may now be used to define the design pattern for each of the four components. Figure~\ref{fig_functional_unit_1} depicts such a pattern of a functional component, while the design patterns of the other components are depicted in Figure~\ref{fig_unit_types} of the Appendix. With these design pattterns modeling a sensor, an actuator, a functional and an appliance one can define all the components in a robotic system that are important for a Self-X module. The next section presents some illustrative examples on how to use such models.

\subsection{Instantiation of a component}
\label{Section_component_design_patterns}
Applying the ontology by implementing it to structure the robot's knowledge base implies that instances of components and their requirements are created. This section presents how design patterns could guide this implementation process. The instantation process follows the 2$^{nd}$ order logic of Section~\ref{background_hypergraphs}.  Two aspects are important in this implementation process. Firstly, in any illustration a white shape represents a \textit{concept} while a grey shape represents an \textit{instance of that concept}. Secondly, assuming that the ontology itself is already implemented to structure the knowledge base, then instances of entities and relations are created with a function $in$ producing a variable, e.g., $c = in(\text{Entity})$, or $r = in(\text{Relation})$, while an attribute is created as a variable \textit{with a value}, e.g., $c = in(\text{Min}) \wedge c = 20.0$. 

\begin{figure}[hb]
    \centering
        \centerline{\includegraphics[width=0.7\columnwidth]{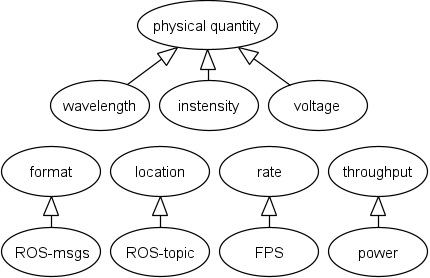}}
        \caption{A further refinement of some properties: \C{ROS-msgs} is a subclass of \C{Format}; \C{ROS-topic} is a subclass of \C{Location}; \C{FPS} (or Frames Per Second) is a subclass of \C{Rate}; \C{Power} is a subclass of \C{Throughput}; \C{Wavelength}, \C{Intensity} and \C{Voltage} are subclasses of \C{Physical Quantity}.}
    \label{fig_subschema}
    \end{figure}

The following subsections illustrate how the pattern of a sensor and a of functional is used to the related parts of a robot's knowledgge base. 
Section~\ref{sec_instance_component} starts by creating the instances of a component and its creations that either own or feature particular properties, followed by Section~\ref{sec_instance_require} where these instances are used to complete the design patterns of a sensor and of a functional with the FR, NFR and ER. Our examples require that \C{RGB Camera} is a subclass of \C{Sensor}, \C{Detector} is a subclass of \C{Functional}, \C{Camera Image} and \C{Robot Pose} are subclasses of \C{Signal} and \C{HealthState} is a subclass of \C{Property}. Figure~\ref{fig_subschema} depicts a further refinement of some properties needed, for example, to support a ROS implementation.

    \begin{figure}[!ht]
    \centering
        \centerline{\includegraphics[width=0.9\columnwidth]{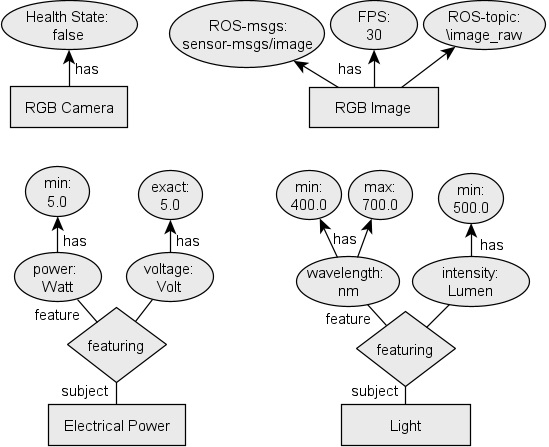}}
        \caption{Four examples to create a first set of instances of the sensor's design pattern. Top left is the instance of an RGB-camera having a health state valued False; Top right is the instance of an RGB-image having the ROS-msgs valued ``sensor-msgs/image'', the ROS-topic valued ``/image\_raw'', the FPS valued 30; Bottom left is an electrical power featuring a minimum power of 5.0 Watt and a voltage of exactly 5.0 Volt; Bottom right is a light featuring a wavelength between 400.0 and 700.0 nm and an intensity of at least 500.0 Lumen.}
    \label{fig_cam_elements}
    \end{figure}

\subsubsection{Instantiation of concepts as component and creation}
\label{sec_instance_component}The sensor that will be defined in the knowledge base is an RGB-camera (component) that produces an RGB-image (creation) as its outcome. It further requires electrical power (resource) and light (physical phenomena) to operate properly. The NFR and ER of the camera will be defined in the next section. We start with the instances of some properties: $m_{1} = in(\text{ROSmsgs}) \wedge m_{1} = \text{``sensor\_msgs/image''}$; $t_{1} = in(\text{ROStopic}) \wedge t_{1} = \text{``/image\_raw''}$; $f_{30} = in(\text{FPS}) \wedge f_{30} = 30$; $q_{0} = in(\text{Quality}) \wedge q_{0} = NaN$. $h_{f} = in(\text{HealthState}) \wedge h_{f} = False$; $m_5 = in(\text{Min}) \wedge m_5 = 5.0$; $m_{400} = in(\text{Min}) \wedge m_{400} = 400.0$; $m_{500} = in(\text{Min}) \wedge m_{500} = 500.0$; $m_{700} = in(\text{Max}) \wedge m_{700} = 700.0$; $x_{5} = in(\text{Exact}) \wedge x_{5} = 5.0$. With those, the next step is to define the instances of the component itself, what it produces and what it requires, see Figure~\ref{fig_cam_elements}, which are formally created as follows:
  \begin{align*}
  &c = in(\text{RGB Camera}),\\
  & \quad c.hasHealthState = \{h_f\};\\
  &d = in(\text{Camera Image}),\\
  &\quad d.hasROSmsgs = \{m_1\}, \quad d.hasFPS = \{f_{30}\},\\
  &\quad d.hasROStopic = \{t_1\}, \quad d.hasQuality = \{q_0\};\\
  &f_1 = in(\text{Featuring});\\
  &e = in(\text{Electrical Power}),\\
  &\quad e.subject = \{f_1\};\\
  &v = in(\text{Voltage}) \wedge v = \text{``volt''}\\
  &\quad v.hasExact = \{x_5\}, \quad v.feature = \{f_1\};\\
  &p = in(\text{Power}) \wedge p = \text{``Watt''},\\
  &\quad p.hasMin = \{m_5\}, \quad p.feature = \{f_1\};\\
  &f_2 = in(\text{Featuring});\\ 
  &l = in\text{Light}),\\
  &\quad l.subject = \{f_2\};\\
  &w = in(\text{Wavelength}) \wedge w = \text{``nm''},\\
  &\quad w.hasMin = \{m_{400}\},\quad w.hasMax = m_{700}\},\\
  &\quad w.feature = \{f_2\};\\
  &i = in(\text{Instensity}) \wedge i = \text{``Lumen''},\\
  &\quad i.hasMin = \{m_{500}\}, \quad i.feature = \{f_2\};
  \end{align*}

The above example will create a first set of instances that are needed to complete the design pattern of a camera in the next section with the emaining FR, NFR and ER relations. Let us explain the 2$^{nd}$ order logic shown above. First, an RGB-camera $c$ is instantiated as a Sensor with a health state that is false, i.e., it cannot be assumed that the camera will produce an output as the system does not yet know whether its requirements are satisfied, it should be infered in a next step. The camera produces a camera image which is instantiated as $d$ having a format and a location as they are typically specified by ROS middleware for cameras. The rate at which the image is produced is 30 frames per second, and since the quality of the image is it not yet known it is valued as being not-a-number ``NaN''. Finally, the creations that are required for the camera to operate properly are instantiated. Since the camera is a sensor it does not require any other data as input. It does need some electrical power $e$ that should be delivered with 5 Watts and at 5 Volts, as is indicated by the featuring relation $f_1$. Similarly, the camera needs some light $l$ that is present in its envrionment which, as indicated by the featuring relation $f_2$, should be in the RGB spectra and thus have a range in wavelength from 400 to 700 nm while having a minimum intensity of 500 Lumen.

\subsubsection{Instantiation of the require relation}
\label{sec_instance_require}
\begin{figure}[!t]
    \centering
        \centerline{\includegraphics[width=0.95\columnwidth]{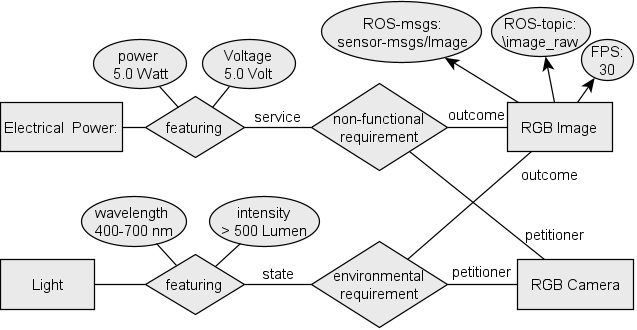}}
        \caption{The instantiation of a complete camera component. Note that attributes of attributes have been summarized, e.g., wavelength 400-700 nm,}
    \label{fig_cam_dp2}
    \end{figure}

The last step is to define the requirements of a component. Each instantiated component should have at least one requirement (i.e. a FR, NFR, ER). The examples presented in the previous section instantiated the component, the creation it produces and the creations it requires. Therefore, next is to link these (prior) instantiations via the instances of the requirement relations specified for that particular type of component. To that extend, recall that the variables $c$, $d$, $f_1$ and $f_2$ have been instantiated in the previous Section~\ref{sec_instance_component}. Then, instantiating the (FR), NFR and ER of the RGB Camera, and thereby completing the design pattern of this camera as is also depicted in Figure~\ref{fig_cam_dp2}, is formally defined as follows:
  \begin{align*}
  & nfr = in(\text{NonFunctional Requirement}),\\
  & \quad nfr.petitioner = \{c\}, \quad nfr.outcome = \{d\},\\
  & \quad nfr.service = \{f_1\};\\
  & er = in(\text{Environmental Requirement}),\\  
  & \quad er.petitioner = \{c\}, \quad er.outcome = \{d\},\\
  & \quad er.state = \{f_2\};
  \end{align*}

\begin{figure}[!t]
    \centering
        \centerline{\includegraphics[width=0.95\columnwidth]{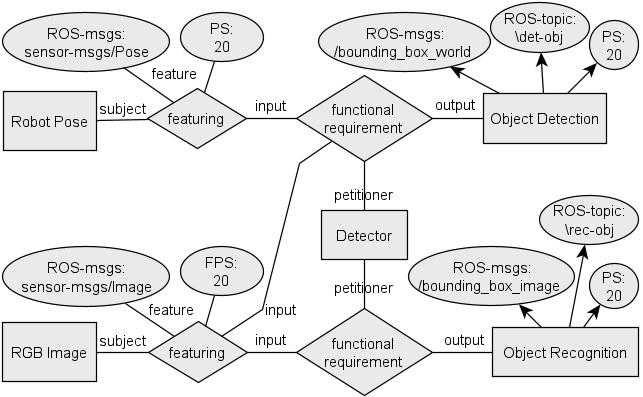}}
        \caption{The instantiation of a complete detector component, where attributes of attributes were summarized. Also, some additional properties were instantiated, such as the ROS-msgs ``bounding-box-image'' and ``bounding-box-world'' (messages defined by YOLO~\cite{bjelonicYolo2018}), the ROS-topics ``rec-obj'' and ``det-obj'' and PS as a subclass of Rate indicating a number per second (in this case 20).}
    \label{fig_det_dp2}
    \end{figure}

When inserting a requirement, one should be aware of all possible combinations and conjunctions. These conjuctions are not present in the complete pattern of the camera model, yet they are present in the pattern of the detector model that is depicted in Figure~\ref{fig_det_dp2} of which its instances may be defined with similar 2$^nd$ order logic as was done for the camera example. The 'or' conjunction of this model is present in the fact that there are two different FRs for which the detector is the petitioner, implying that the detector may be set to either one of the possible outputs it can create. A first output is a \C{Recognized Object} (subclass of \C{Information}). This output is a list of objects that are recognized by the detector in the camera image, which is why the FR linked to this output shows that only a camera image is required as input signal. A second output of the detector is \C{Detected Object} (subclass of information). This output is a list of all objects that are recognized in the camera image \textit{including an estimated position of the object in the real world}. For that reason, the FR linked to this output shows that both the camera image `and' the robot's pose are required, since without the position of the robot in the real world the detector is not able to estimate the position of the object in the real world from its position in the image. So, via this type of specification, one can express which input is needed for which output option of the detector. Further note that the output data has a location, i.e., a ROS-topic, while the input data does not. This is because it has not yet been inferred which actual data that is present in the system satisfies the FRs of the detector and where such data is located.

\section{Inferring a configuration and modeling behaviors}
\label{Section_behavior} 
The previous section introduced the design pattern of a single component with some illustrative instances of a particular sensor and a functional, i.e., the RGB camera and the detector. This section continues with the design pattern of the system configuration that is necessary to execute a behavior. Recall, from Figure~\ref{fig_general_setup}, that each behavior is the result of a particular set of components, with input-output relations that define the system configuration for that behavior. Therefore, creating a design pattern for a configuration of components per behaviour means to design the knowledge structure between component and behaviour via its configuration. To start, some rules are introduced that model an understanding of which components can actually take part in a configuration as all their requirements are met. This understanding will be then be used to define the configuration per behavior as a design pattern.

\subsection{Realizing the requirements for actual processing}
\label{Section_realizing_processing_ontology} 
Each component may have three different requirements that need to be fulfilled, i.e., a functional, a non-functional and an environmental requirement. In case all requirements of a component are realized, for example by other components or by the environment, then the component is said to have a complete proces and produces its outputting creation. Therefore, let us introduce \C{Realizing} and \C{Processing} as subclasses of \C{Relation}.

\begin{figure}[!ht]
     \centering
     \includegraphics[width=0.99\columnwidth]{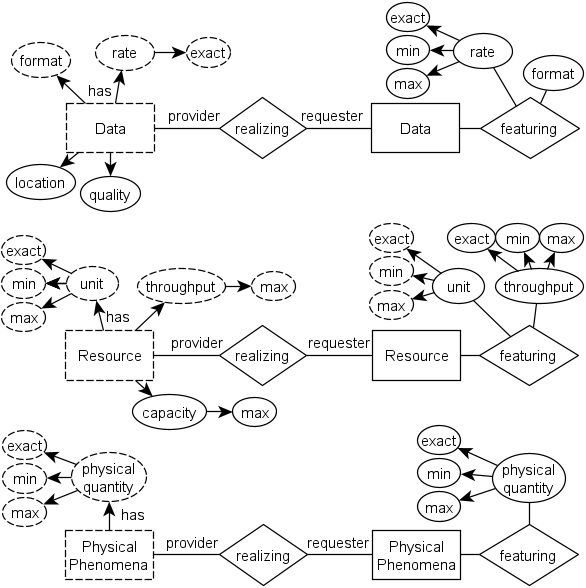}
     \caption{Three design patterns of \C{Realizing} as a subclass of \C{Relation} taking place between two of the same creations. Note that \C{Range} has already been substituted with its possible subclasses of \C{Exact}, \C{Min} and \C{Max}, though for data it is assumed that there is exactly one rate at which it can be provided.}
     \label{fig_realizing}
  \end{figure}

Let us start with the design patterns of \C{Realizing} depicted in figure~\ref{fig_realizing}. The realizing relation links two creations of the same type, i.e, data with data, a resource with a resource and a physical phenomena with a physical phenomena. One creation in this relation plays the role of \Li{Provider} that \textit{has} a set of properties while the other creation plays the role of \Li{Requester} that \textit{feateres} the same type of properties. Rules in 2$^{nd}$ order logic infer whether a realizing relation should be instantiated between two creations, i.e., when the properties of the requested creation are met by the creation that is provided.

There are three types of creations relevant to \C{Realizing}, which are data, resources and a physical phenomena. When the creations concern data, then a realizing relation is implied if: ($i$) the format as featured by the requested data is equal to the format of the data provided and; ($ii$) constraints on the rate as \textit{featured} by the requested data are not in violation with that what the providing data \textit{has} to offer as update rate. The  2$^{nd}$ order logic to define the inferrence rule of a realizing relation between two instances of data is presented hereafter. Note that it may stil be the case that multiple instances of data are available in the knowledge base that comply with this rule. The selection of which instance will provide the actual data for a component is defined later when discussing a behaviour. Further, note that quality is not checked in this logic. This is because quality is a numerical variable that (in future research) will take part in an equation, and not in a constraint: an equation that determines a quality of the data that is outputted by a component based on the quality of that the data that is provided to this component. Also, note that when a realizing relation is instantiated, then the component that required such data will now also have the location at which it can retrieve that data. 
 \begin{align*}
& D_r  = \, \{d \in Data \, | \, d.subject \in Featuring \}.\\
&\forall d_r \in D_r: \\
& \quad F_t = \{f_t \in \text{Featuring} \, | \, f_t \in d_r.subject \},\\
& \quad F_o  = \{f_o \in \text{Format} \, | \, f_o \in f_t.feature \wedge f_t \in F_t\},\\
& \quad R_t  = \{r \in \text{Rate} \, | \, r \in f_t.feature \wedge f_t \in F_t\},\\
& \quad E_x  = \{e \in \text{Exact} \, | \, e \in r_t.hasExact \wedge r_t \in R_t\}, \\
& \quad M_i  = \{m \in \text{Min} \, | \, m \in r_t.hasMin \wedge r_t \in R_t\},\\
& \quad M_a  = \{m \in \text{Max} \, | \, m \in r_t.hasMax \wedge r_t \in R_t\}, \\
& \quad D_p  = \{d \in \text{Data} \, | \, d \not\in f_t.subject, \forall f_t \in F_t\}.\\\\
& \quad \exists d_p \in D_p \, | \, \big(d_p.hasFormat \cap F_o \neq \emptyset \big) \wedge \\
& \quad\quad\quad \big( (d_p.hasFormat.hasExact \cap E_x \neq \emptyset) \vee \\
& \quad\quad\quad (\exists e_p \in d_p.hasFormat.hasExact \wedge \exists m_i \in M_i | e_p \geq m_i) \vee \\
& \quad\quad\quad (\exists e_p \in d_p.hasFormat.hasExact \wedge \exists m_a \in M_a | e_p \leq m_a) \big) \\
& \quad \rightarrow r \in Realizing,\, r.requester = \{d_r\}, \, r.provider = \{d_p\}.
  \end{align*}

The implication of a \C{Realizing} for the other two creations, i.e., resources and physical phenomena, is inferred with rules that are similar to the one shown above for data. When the creations concern resources, then an actual realizing relation should be implied if: ($i$) the unit and throughput as \textit{featured} by the requested resource are of the same type as that of what the providing resource \textit{has} to offer and; ($ii$) the constraints on the amount of unit and throughput as \textit{featured} by the requested resource are not in violation with that of what the providing resource \textit{has} to offer. Interestingly, it also allows the system to keep track of each providing resource, for example on how much of its throughput is still available, or on the time that this resource will remain avialable when taking into account the capacity of this providing resource (remaining time equals capacity divided by throughput). When the creations concern physical phenomena, then such a realizing relation should be instantiated if the physical quantity as \textit{featured} by the requested phenomena is of the same type as what the providing phenomena \textit{has} to offer, and that the constraints on the amount of this physical quanity as is \textit{featured} by the phenomena are not violated by what the providing phenomena \textit{has} to offer. Note that, as physical phenomena are often provided by the environment, they should be instantiated in the knowledge base when present.\\

  \begin{figure*}[t]
     \centering
     \includegraphics[width=0.9\textwidth]{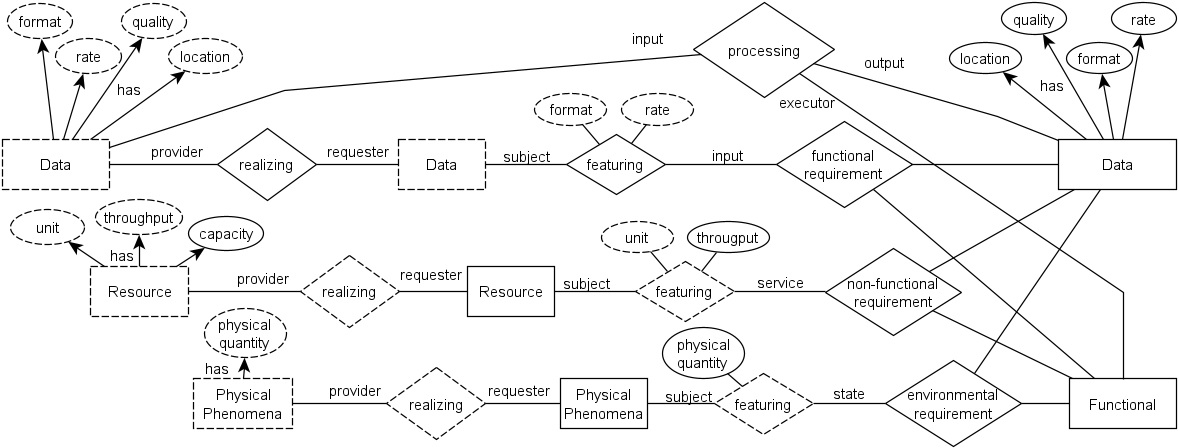}
     \caption{Design pattern of \C{Processing} as a subclass of \C{Relation}. In this case the processing relation in defined for a functional component, while the processing relation itself should only be instantated when all the requirements of thel component are realized (with the realizing relation), such as data or services within the system itself, or states in the environment of the robotic system.}
     \label{fig_processing}
  \end{figure*} 

Now that realizing relations are modeled, let us continue with the design pattern of \C{Processing} depicted in figure~\ref{fig_processing}. The processing relation models a summary of a complete component in operation, i.e., the relation specifies the \Li{Input} data (if required) of a component with its \Li{Output} data (if produced). The component itself plays the role of \Li{Executor}. Resources and physical phenomena cannot play the role of input nor output, implying that an appliance will never be linked to a processing relation. This is because the ontology models how a behavior of a robotic system depends on the performance of its components, in which only the performance of sensors, actuators and functional components are considered and it is assumed that any nonfunctional requirements of such components are sufficiently met. The summarizing aspect of the processing relation is that it bypasses certain properties of the data, resources and physical phenomena, as those were already checked by inference rules upon instantiating the necessary realizing relations. Hence, the processing relation may only be implied when all requirements of a component have been met, as formalized in the 2$^{nd}$ order logic presented below. Note that this logic checks per possible product (or output) of the component whether all of its requirements (FR, NFR and ER) are met.
 \begin{align*}
&C_p = \, \{c \in \text{Sensor} \cup \text{Actuator} \cup \text{Functional} \}. \\
&\forall c_p \in C_p: \\
& \quad D_o = \{d \in \text{Data} \, | \, d \in r.product \wedge r \in c_p.petitioner\} \\
& \quad \forall d_o \in D_o: \\
& \quad\quad D_i  = \{d \in Data \, | \, f \in d.subject \wedge \\
& \quad\quad\quad\quad\quad\quad\quad\quad r_{fr} \in f.input \wedge r_{fr} \in d_o.output \},\\
& \quad\quad S_i  = \{s \in \text{Resources} \, | \, f \in d.subject \wedge \\
& \quad\quad\quad\quad\quad\quad\quad\quad r_{nfr} \in f.service \wedge r_{nfr} \in d_o.outcome \},\\
& \quad\quad P_i  = \{p \in \text{Physical Phenomena} \, | \, f \in d.subject \wedge \\
& \quad\quad\quad\quad\quad\quad\quad\quad r_{er} \in f.state \wedge r_{er} \in d_o.outcome \}.\\\\
& \quad\quad \exists d_i.requester (\forall d_i \in D_i ) \,\, \wedge \\
& \quad\quad \exists s_i.requester (\forall s_i \in S_i ) \,\,\, \wedge \\
& \quad\quad \exists p_i.requester (\forall p_i \in P_i ) \\
& \quad\quad \rightarrow p \in Processing,\,  p.input = D_i, \, p.output = \{d_o\}.
  \end{align*}

The above logic inserts infers a new instance of \C{Processing}, if for the FR, NFR and ER linked to the output/outcome of a component it holds that:

($i$) all data linked to the FR via a featuring relation, in which they play the role of input, are also linked to a realizing instance, which exists iff another data exists that \textit{has} the properties of the requested data.

($ii$) all resources linked to the NFR via a featuring relation, in which they play the role of service, are also linked to a realizing instance (which exists iff another resource exists ithat \textit{has} the properties of the requested resource).

($iii$) all physical phenomena linked to the ER via a featuring relation, in which they play the role of state, are also linked to a realizing instance (which exists iff another physical phenomena exists ithat \textit{has} the properties of the requested phenomena).\\

\begin{figure}[!ht]
     \centering
     \includegraphics[width=0.99\columnwidth]{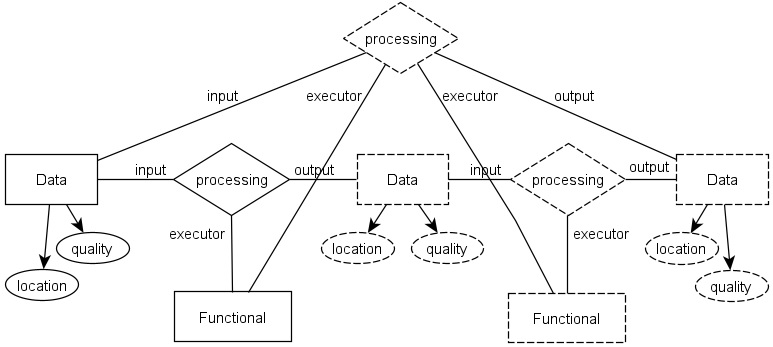}
     \caption{Design pattern of \C{Processing} in the case of a transitivity.}
     \label{fig_processing_transitivity}
  \end{figure}

Further, it is important to note that the processing relation is a transitive relation, as is depicted in Figure~\ref{fig_processing_transitivity}. Such a transitive relation is defined with a transitivity rule stating that if the output creation of one component is used as the input creation of another component, and each component plays the role of executor in its own processing relation, then a larger processing relation can be defined linking both two components as executor, and with its input is linked to the input of the first component while its output is linked to the output of the second component. This rule applies to any number of components that share input-output connections.

\subsection{Defining a behaviour of the robotic system}
\label{Section_ontology_behavior}
\begin{figure}[!ht]
     \centering
     \includegraphics[width=0.95\columnwidth]{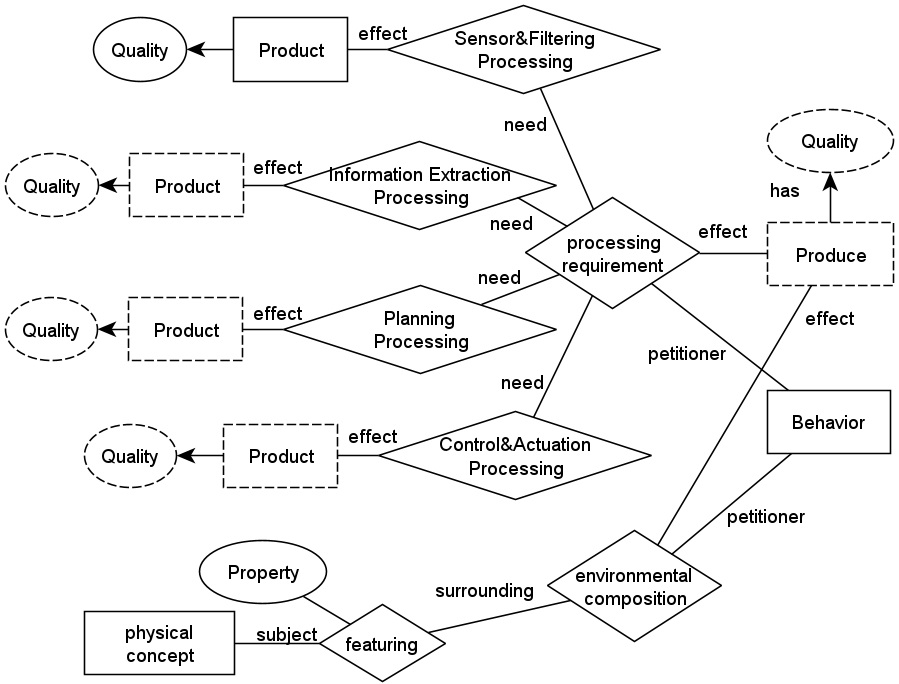}
     \caption{Ontology model of a Behavior (inputs Commodity features a Quality, output Commodity has a Quality as MoP).}
     \label{fig_behavior}
  \end{figure}

The purpose of Self-X is to assess how well the robot is performing a task or behavior. Discussions on what is a task or a behavior are not yet settled, see [ref], though there is an agreement that a behavior will be part of a task. This also means that, in order for a robot to execute a behavior it would require a particular system configuration producing such desired behavior. Therefore, this section will describe the connection between the ontology on the system components and the system configuration, as introduced in the previous Section~\ref{Section_CDROM_Concept}, and the concept of a Behavior. In view of the use-case of this article an example of a robot's Behavior is to search a room to find victims. A robot can execute this behavior either visually or acoustically, each requiring their own system configuration as depicted in Figure~\ref{fig_search_visual_acoustic}. It was already denoted that the system configuration is divided into four different subsystems (Sensing\&Filtering, Information Extraction, Planning and Control\&Actuating) and that the $processing$ relation is able to capture each partial configuration that the system would be able to invoke under the current (or planned) conditions. Therefore, a system engineer defines a Behavior by stating which results (or Commodities) are required per configuration (or $processing$ relation) of the subsystem. And as the $processing$ relation already checks Format, Throughput and Capacity of the FRs and the NFRs per Unit, note that the dependency of the Commoditie's Quality and on the Environmental Composition remains. This is also intended, as the actual Quality that the Behavior accomplishes is a function of the environment in which it is operating its Behavior and of the Quality per subsystem configuration. An illiistration of the ontology model for a Behavior is depicted in Figure~\ref{fig_behavior}.

\subsection{Modelling of configurations and behaviors}
\label{Section_behavior_ontology} 
The models that are described in previous sections are used to guide the implementation process for components, configurations and behaviors. Section \ref{Section_component_design_patterns} describes the instantiation of components using these models, in this section the instantiation of configurations and behaviors is described and will be illustrated via examples. 

\subsubsection{Instantiation of configurations}
In the current model, the instantiation of configuration is infered via logical rules that infer the $realizing$ and based on that, the $processing$ relation. The realizing relation rule checks if there is any real world concept that fulfills an input requirement of a component and if so, insert the relation. The rule can be specified as follows:
\\
$ X \subset (x \in creation|x.hasProperty: p1) \wedge \\
Y \subset (y  \in creation|y.hasProperty: p2| \langle y, p2 \rangle : featuring) \wedge \\
Y \neq X \wedge \\
Y \in X \wedge \\
p1 \in p2  \\
\to insert(\langle X, Y \rangle : realizing) $

\begin{figure*}[!ht]
	\centering
	\includegraphics[width=0.99\textwidth]{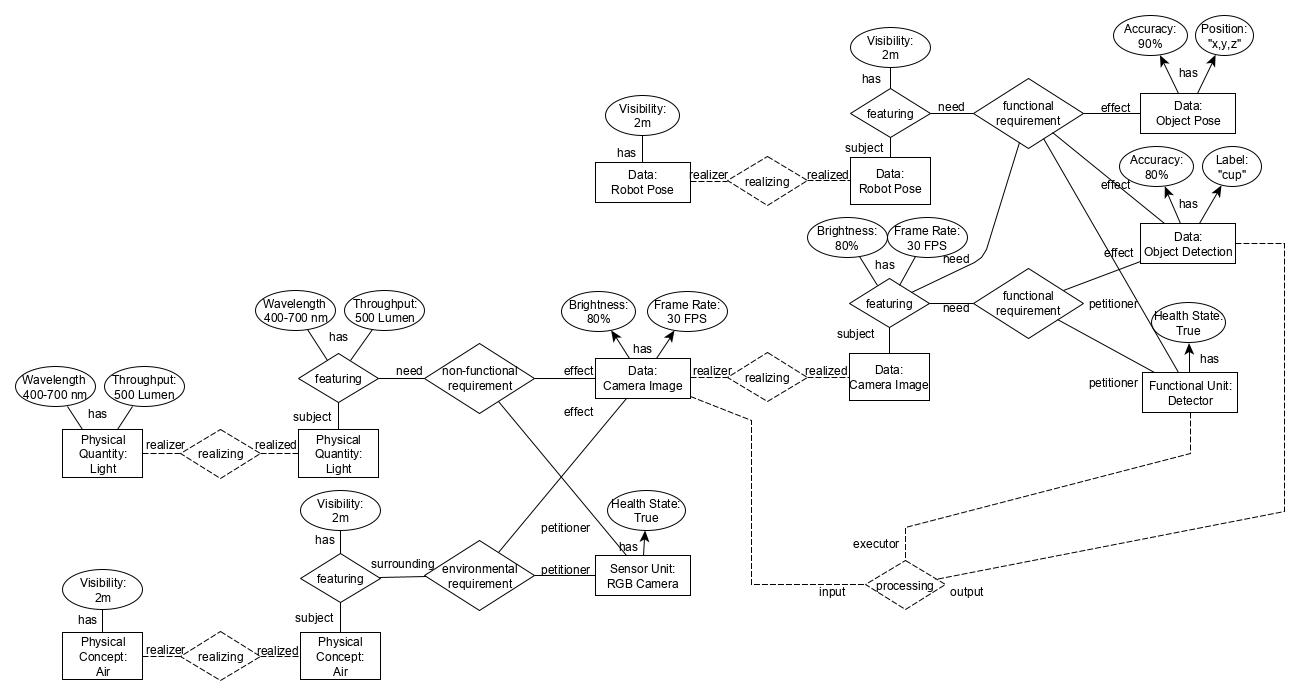}
	\caption{Instantiation of the realizing rule (dotted lines means that the relation is inferred)}
	\label{fig_DPrealizing}
\end{figure*}

An example of the realizing relation can be found between the camera output and the detector input requirement. Lets assume that the $detector$ component has an input requirement for a $camera$ $image$ with a $rate$ of $30FPS$. The realizing rule checks if there is a concept instantiated that adheres to this requirement. In this example, the rule will find that the output of a $RGB$ $camera$ realizes the requirment since it outputs an $camera$ $image$ with a $rate$ of $30FPS$. Consequently, the realizing rule automaticlly generates a $realizing$ $relation$ between the real $camera$ $image$ and the input requirements to indicate that the one realizes the other (see Figure \ref{fig_DPrealizing}). This automatic process of matching input requirement and instantiated real world data will be done for all specified requirements via the realizing rule as descriped in Section \ref{Section_realizing_processing_ontology}.

The possible configurations become more clear via the linking of components by the realizing relation. In the previous example the realizing relation linked the camera and detector as the ones output realizes the others input and so can be used in a configuration. The configuration itself is made more explicit via the processing relation. This relation is logically infered and links the required initial input(s), to a final output via all intermediate unites that are needed in that process. The rule can be specified as follows:
\\
$\{I,Y,F,R,O,C\} \subset \{\{\{i \in creation\},\{y \in creation\}, \\
\{f \in featuring\}, \{r \in realizing\},  \\
\{o \in creation\}, \{c \in component|c.hasHealthState = "True"\}\}|  \\
\langle f, c, o \rangle : requirement, \\
\langle y, f \rangle : featuring \} \\
\langle i, y \rangle : realizing\} \\
\to insert(\langle i, c, o \rangle : processing) $

\begin{figure}[!b]
	\centering
	\includegraphics[width=0.95\columnwidth]{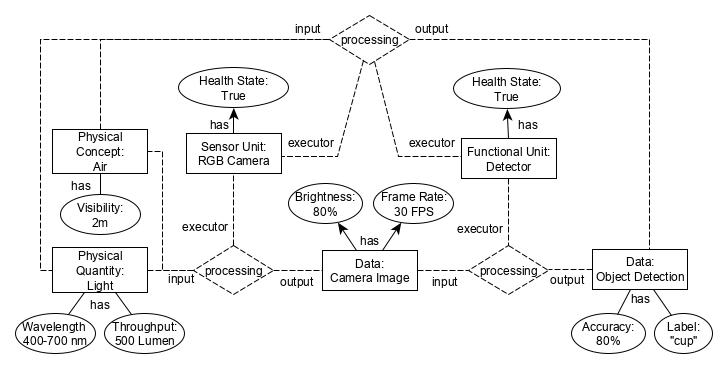}
	\caption{Instantiation of the processing rule (dotted lines means that the relation is inferred)}
	\label{fig_DPprocessing}
\end{figure}

The processing relation of the previous example is shown in Figure \ref{fig_DPprocessing}. In total, three processing relations are inferred. The first one links the physical inputs $air$ and $light$, via the executor $rgb camera$ to the $camera$ $image$ output. The second links the $camera$ $image$ as input for executor $detector$ to process it to the output $object$ $detection$. Just like the pervious relation, the automatic process of matching input, output and components via a processing relation will be executed for all realizing relations. Furthermore, the rule is transitive, which means that if there are two processing relations as just desribed, a new relation will automatically be generated between input: $air$, component: $RGB$ $camera$, component: $detector$ and output: $object$ $detection$ which is the thirth relation shown in Figure \ref{fig_DPprocessing}.

Since the configurations are infered via logical rules, no extra instantiation is needed for these relation. However, if it is desired to specify (sub-) configurations, this can be done by specifying different types of processing. One could for example specify a sensor\&filtering processing with output: $object$ $detection$ with $Accuracy$ $0.9$. Logical rules can be used to infer if there exists a healthy configuration that outputs an $object$ $detection$ with the desired accuracy and so, if there exists a (sub-)configuration for the desired sensor\&filtering processing.

\subsubsection{Instantiation of behaviors}
To reason about actions, one should instantiate how these actions can be executed, which we call behaviors. These behaviors are then linked via the inferred processing relation to the configurations that fulfill them. An example of a behavior is a 'person detection via camera' or a 'person detection via speech' both describing the exectution of the 'detection' action. Furthermore, a behavior requires an effect; e.g. both search behavior might result in a detection. In the ontology, one should specify the possible behaviors and link them to an effect and the related action. This instantiation looks as follows:
\\
$insert \\
(b \in behaviour, b.hasName = [name]); \\
(d \in Data, d.hasProperty: [property]); \\
pc (\{petitioner: b, effect: d\}: processing-requirement);$

\begin{figure}[ht!]
	\centering
	\includegraphics[width=0.8\columnwidth]{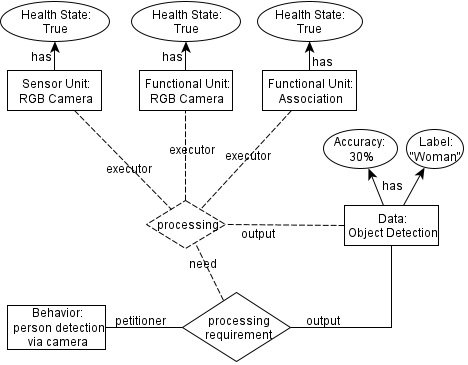}
	\caption{Instantiation of the visual configuration}
	\label{fig_visual_configuration}
\end{figure}

\begin{figure}[ht!]
	\centering
	\includegraphics[width=0.7\columnwidth]{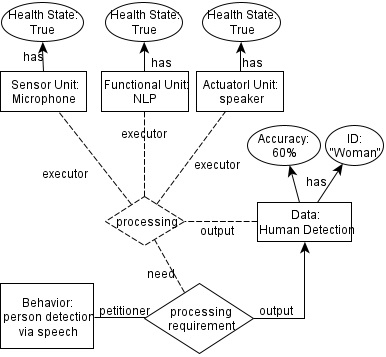}
	\caption{Instantiation of the acoustic configuration}
	\label{fig_acoustic_configuration}
\end{figure}

If the processing requirement is instantiated, logical reasoning rules can infer which processing relations are required for the execution of the behavior. If multiple behaviors can be chosen to execute an action, performance might be a factor to take into account in the selection. 
In the running example, the configurations and environment that the AS has to operate in is used to predict the performence. In the example the system could chose between a 'person detection via camera' and a 'person detection via speech', Figure \ref{fig_visual_configuration} and \ref{fig_acoustic_configuration} show the instantiations of the possible configurations per behavior. In this case, the visibability of the air is very limited (0.5) as is the light quality and as they are inputs for the 'person detection via camera' behavior, the effect will have a low quality as well (e.g. 30 percent accuracy). The effect of the 'person detection via speech' behavior has an accuracy of 60 percent since it is not affected by any unfavorable environmental conditions. In the case that the AS tries to optimize the quality of the output, it would choose to detect a persons via speech.

\section{Real-life example}
\label{Section_RealLife_Example}
\begin{figure}[!ht]
     \centering
     \includegraphics[width=0.75\columnwidth]{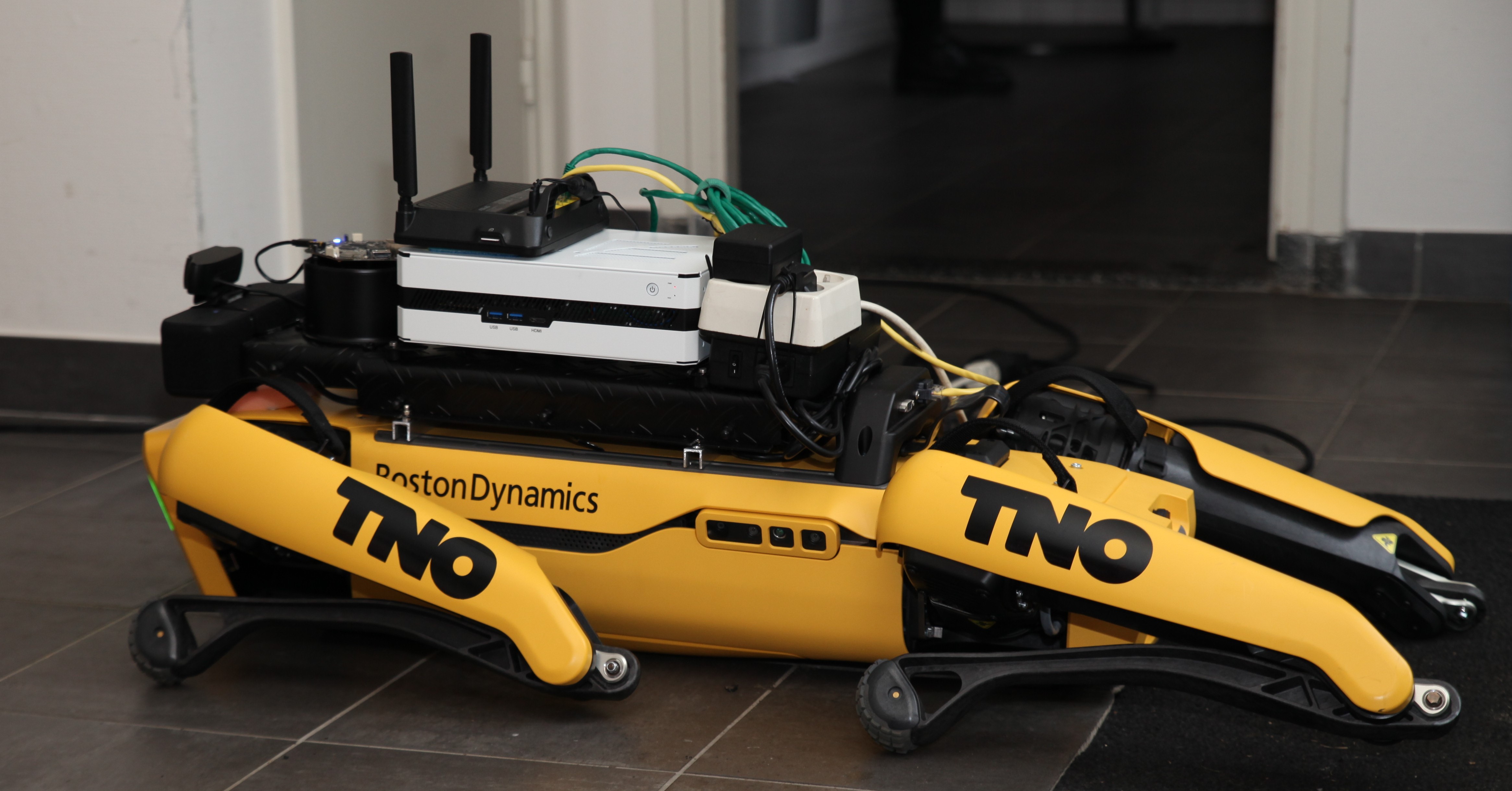}
     \caption{The robotic system used for SelfX.}
     \label{fig_spot_robot}
  \end{figure}

The Self-X approach has been implemented on a robotic system to assess the expected performance of some of its most important behaviors before creating a plan. The hardware integration involved the SPOT robot from Boston Dynamics~\cite{spot}, a USB camera from Logitech and the ReSpeaker from Seeed Studio~\cite{respeaker} with a build-in microphone array and a connection to a Bluetooth speaker. All this hardware was digitally integrated via drivers with the ROS2 environment running on the embedded PC that is mounted on the back of SPOT (see Figure~\ref{fig_spot_robot})
  \begin{figure*}[b]
     \centering
     \includegraphics[width=1.0\textwidth]{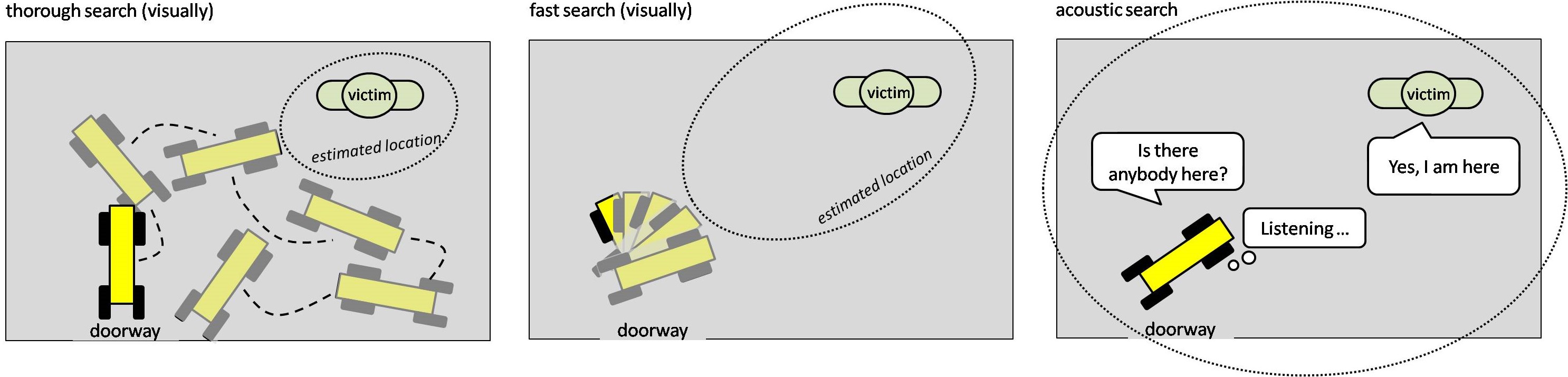}
     \caption{The three different search behaviors of the robot, each resulting in a particular accuracy of the victim's location.}
     \label{fig_search_beaviors}
  \end{figure*} 

\subsection{Task and behaviors}
The robot is tasked to locate people in a residential house by searching the different rooms in this house. To do so, the robot has one GoTo behavior by which it is able to go from one room to another and three types of Search behaviors by which it is able to search a room and locate victims. Once the robot entered a room in will plan one of its Search behaviors. To decide which Search behavior is best for the room, the performance of each behavior is predicted by SelfX based on the availability and the quality of internal components (ROS2 nodes) and based on environmental conditions (light conditions, noise conditions and room size). The three different Search behaviors that were implemented on the robot are depicted in figure~\ref{fig_search_beaviors} and yield:
  \begin{itemize} 
	\item Thorough search: The robot will plan and walk to different waypoints in a room. At each waypoint the camera image is analyzed on possible victims. Since the robot visits various waypoints one can expect that a victim is detected multiple times, and also at close proximity, which further implies that the association component (associating multiple detection to the same victim) is able to improve the accuracy with which victims are located;
	\item Fast search: The robot will stay put in the room and look at different viewpoints by moving its shoulders. At each viewpoint the camera image is analyzed on possible victims. Since the robot quickly scans the room one should expect that a victim is detected ones, possibly at larger distance, which further implies that the accuracy with which persons are located is less compared to a thorough search;
	\item Acoustic search: The robot will stay put in the room and face the center of the room. Then it will shout: ``Is there anybody here?'' and take audio recordings for several seconds after it posed the question. It will repeat this several times. These recording are then processed and when an English reply is heard it will annote a detection somewhere in that room. The accuracy of the estimated location will therefore depend on the size of the room (its diagonal in the floor map).
  \end{itemize}

\subsection{Configurations per behavior}
The robot will invoke a particular configuration for each of the three Search behaviors. To keep the illustrative case-study of SelfX concise this real-life example only assesses (and shows) the Monitor and the Analyze configurations of the robot (recall the MAPE-K architecture in Section~\ref{Section_Related_Work_System_Description}). Figure~\ref{fig_example_selfx} depicts two Monitor configurations and two Analyze configurations: one Monitor and one Analyse configuration for conducting an acoustic search behavior; and one Monitor and one Analyse configuration for conducting either the thorough or the fast search behavior. Note that the sensor data for both the thorough and fast search is processed similarly, as their difference is in the Plan and Execute configurations which are not addressed in this illustrative case-study. Because of that, there is also a difference in the \C{MeasureOfPerformance} of the product that is created by the Search behavior, i.e., the product of a list of victims and their location in the room. More details on this difference is presented in the next section.

Here, let us continue with the knowledge structure, or knowledge hypergraph, that results after applying the developed ontology on these configurations and behaviors. At a higher level each behavior is linked to its configuration via the concept called \C{ProcessingRequirement}. This requirement collects all the components that need to be up and running for the behavior, while more details on their actual configuration can be found by querying the realizing relations between its components.
\begin{itemize} 
	\item The processing requirement of the acoustic search is depicted in the knowlege graph of figure~\ref{fig_acoustic_configuration}. Herein, the microphone will be triggered to start recording for 3 seconds after the speaker produced the question ``Is there anybody here?''. This recording will be turned into a text sentence by the speech recognition component, which produces an empty text in case no reply is recorded. Finally, some natural language processing is applied on the text sentence to detect whether it is a human reply, i.e., a spoken language. In the case of such a detection this last component will produce a newly detected victim having a position somewhere in that room. Its instances of the microphone, speech recognition and natural language processing are depicted in figure XXX along with the instances of the resources and phyiscal quantities that these components require. The actual value of physical quantities is estimated by the configuration assessment (see also figure~\ref{fig_example_selfx}) and is further described in the next section;
	\item The processing requirement of the thorough search and of the fast search is depicted in the knowledge structure of figure~\ref{fig_visual_configuration}. Herein, the camera will be triggered to take an image once the robot is at its pre-planned waypoint, or viewpoint. The image is further processed by an object detection component that will detect if a human is present in that image and what the bounding box of that human in the image is. Finally, the object association component converts each bounding box into an estimated position of a human in the room, using the estimated pose of the robot. It further decides whether the detections correspond to newly detected victims or whether it should be associated to victims that were already known to be in that room, for example from prior detections. In the case of a such a detection, then this final component of object association will produce a newly detected victim with an estimated position. Its instances of the camera and of the object detector are depicted in figures~\ref{fig_cam_dp2} and~\ref{fig_det_dp2}, repsectively, while its instance of object association is depicted in figure XXX. Instances of the resources and physical quantities that these components may require are illustrated in figure XXX.
  \end{itemize}
The actual implementation of this hypergraph as a knowledge base was done in GRAKN~\cite{Grakn_blog}, which allows to implement the ontology, the instances and the 2$^{nd}$ order logic inserting the realizing and processing relations as inference rules. In addition, the knowledge base was extended with an ontology and instances to model the floor plan of the house, as well as the location and position of any living beings and artifacts, for which the interested reader is refered to~\cite{Sijs_iros}. This later part of the knowledge base is used to capture conditions that are relevant for assessing a configuration, such as the size of a room.
  \begin{figure}[!ht]
     \centering
     \includegraphics[width=0.95\columnwidth]{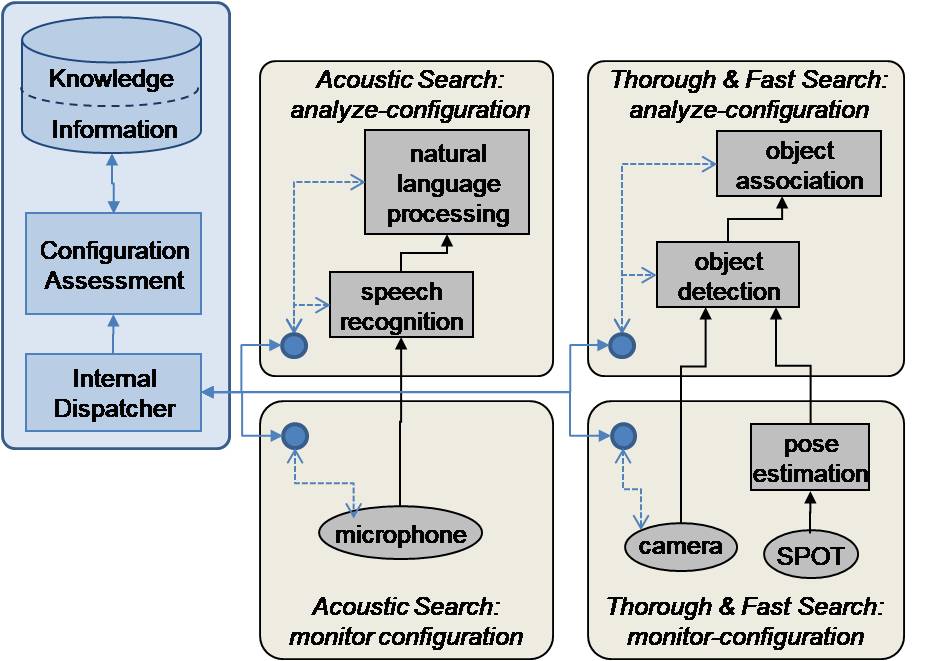}
     \caption{The two types of configuration of which one will be invoked in case the robot will conduct one of its three Search behaviors.}
     \label{fig_example_selfx}
  \end{figure} 

\subsection{Configuration assessment}
The robot is assumed to be in a room before it decides which of the three search actions to perform. This decision depends on the availability of the components that are required for a configuration and on their expected performance. This section presents how this performance is predicted: by matching the current internal readings on conditions and qualities to the prior experiences of the robot. Therefore, let us start by introducing the internal readings one the quality of created data and on the actual values of the physical quantities that are relevant. 

\textit{Microphone}: The quality of the audio-file that is produced by the microphone is measured by the amount of acoustic, background noise ($N$ in dB) versus the expected intensity of a human voice ($S \approx 70$ dB). Suitable implementations that estimate the level of this background noise based on the microphone signals can be found in open source communities. 

\textit{Speech recognition}: The quality of the text sentence produced by the speech recognition is measured by the probability that the sentence was spoken by a human, i.e., the probability of detecting a human ($P(human)$). To compute this probability a Python package called ``langdetect''  was used that takes a text sentence as input and returns a value in between 0 and 1 on how certain that sentence is of a particular language.

\textit{Natural language processing}: The quality of the newly detected victims produced by the natural language processor is measured by the position inaccuracy with which the victims are located ($D$ in meters). This inaccuracy $D$ is set equal to half of the room's diagonal.

\textit{Camera}: The quality of the RGB-image that is produced by the camera is measured by its values for brightness ($B$), contrast ($C$) and Brisque features ($b_{0}$, $b_{1}$, ... $b_{16}$) \cite{brisque}, for which several implementations can be found in open source communities. Further, the camera requires particular light conditions, which is an intensity of at least 50 Lumen within the visible spectrum of 400 - 700 nm).

\textit{Object detection}: The quality of the (human) detections of the object detector is measured with probility that the detection also corresponds to a human ($P(human)$). In this case this probability was set to the confidence that such detectors typically produce for each detection, even though we are aware of the biases in these confidences.

\textit{Object association}: The quality of the newly detected victims produced by the object associator is measured by the position inaccuracy with which the victim is located ($D$ in meters). This inaccuracy $D$ is a combination of the robot's position accuracy ($\delta \sim 0.25$ m) and the distance $d$ between the robot and the detected victim (typically a camera will produce less position accuracy at larger distances). More precisely, based on some heuristics the position inaccuracy is: $D = \delta + \sqrt{d}$.

The next step is to predict the \C{MeasureOfPerformance} of all search behaviors based on the actual conditions and on the quality and availability of components. Assuming that the robot is in the room that it wants to search, then the conditions in that room and the qualities acquired as internal readings are:
  \begin{itemize} 
	\item Acoustic search: The background noise $S$ and the position inaccuracy $D$, where $D$ is obtained from the room-diagonal that is retrieved from the size of the room;
	\item Fast search: The brightness $B$, contrast $C$, Brisque features $b_0$, ..., $b_{20}$ and position inaccuracy $D = \delta + \sqrt{d}$, where the distance $d$ is the expected distance between the robot and a possible human (approximated as the -average- distance from the robot to the opposite wall). Further, light conditions, being the intensity in the visible spectrum, are acquired as internal reading from a seperate sensor (foto resistor).
	\item Thorough search: The brightness $B$, contrast $C$, Brisque features $b_0$, ..., $b_{20}$ and position inaccuracy $D = \delta + \sqrt{d}$, where the distance $d$ is the expected distance between the robot and a possible human (approximated as the -average- distance between the waypoints that have been planned when searching the room thoroughly). Further, light conditions, being the intensity in the visible spectrum, are acquired as internal reading from a seperate sensor (foto resistor).
  \end{itemize}

Finally, these internal readings are used to predict the performance of a search behavior, independent on whether the search is acoustic, fast or thorough. This performance of a search behavior is measured by two variables: 1. the probability that if a human is present in the room, then it is also detected, i.e., $P(human)$, and 2. the position inaccuracy with which a human is detected. Since the latter measure is already available, the only measure that still needs to be predicted is $P(human)$. This prediction is done with a data-driven approach, i.e., learned from experimental data. Hereto, 25 experiments were conducted, accross different rooms and varieties of human presence. The robot was tasked to search for the human thoroughly, fastly or acoustically. While searching, the above internal readings were recorded as well, thereby producing a large data set of conditions, of qualities and of results on whether the human was actually detected. The data related to the conditions and the qualities was then clustered using a clustering appproach known as self-organizing-maps [ref]. Such a map was created for each of the three behaviors. Each $i^{th}$ point in a cluster would, idealy, have similar conditions and quality measures compared to other points in that cluster, while it would also have a Boulean result on whether the human was detected correctly or not, i.e., ($P_{i}(human)$ is either 0 or 1). Then, the measure of performance of that search behavior under the conditions and qualities of that cluster, is computed by averaging over $P_{i}(human)$ for all points $i$ in that cluster. The collection of clusters then creates a so called map that is used to make a prediction on the measure of performance $P(human)$ when conducting the type of search behavior that relates to the map. This prediciton is done by taking the current conditions and qualities and use those for finding the best-matching-unit (point) in the map, and its cluster, thereby obtaining a prediction of $P(human)$ that corresponds to that cluster. 

insert figure

This completes the implementation of the Self-X ontology and how it is used in combination with online, internal readings for assessing a configuraiton and predicting the performance of a behavior. The next section will show some results of practical experiments in which this self-asessment is used to decide which type of search is most effective when entering a room given the components that are available and the condititions of that room.

\section{Experimental results}
\label{Section_Experimental_Results} 
dedductive reasoning: realizing and processing relations
dedcutive reasoning: the room has these conditions (light and size), which onfiguration allows me to have so much accurate detections?
indcutive reasoning: spot has these and these meters from its origin, its origin was in the center of the hall, therefore spot is in the living
inductive reasoning: the performance of my visual seach was bad.Let me check with an acoustic one whether no-one was here. Store this probability of success in my knowledge-base (in this context with these conditions I did not found the victim with my fast-search but with my acoustic search)
indictuve reasoning: the visual search failed for unknown reasons in the prior two rooms. Therefore I should maybe start with using the acoustic approach
  \begin{figure}[!ht]
     \centering
     \includegraphics[width=0.95\columnwidth]{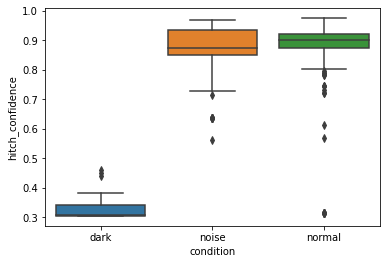}
     \caption{Boxplot hitch confidence.}
     \label{fig_example_selfx}
  \end{figure} 
  \begin{figure}[!ht]
     \centering
     \includegraphics[width=0.95\columnwidth]{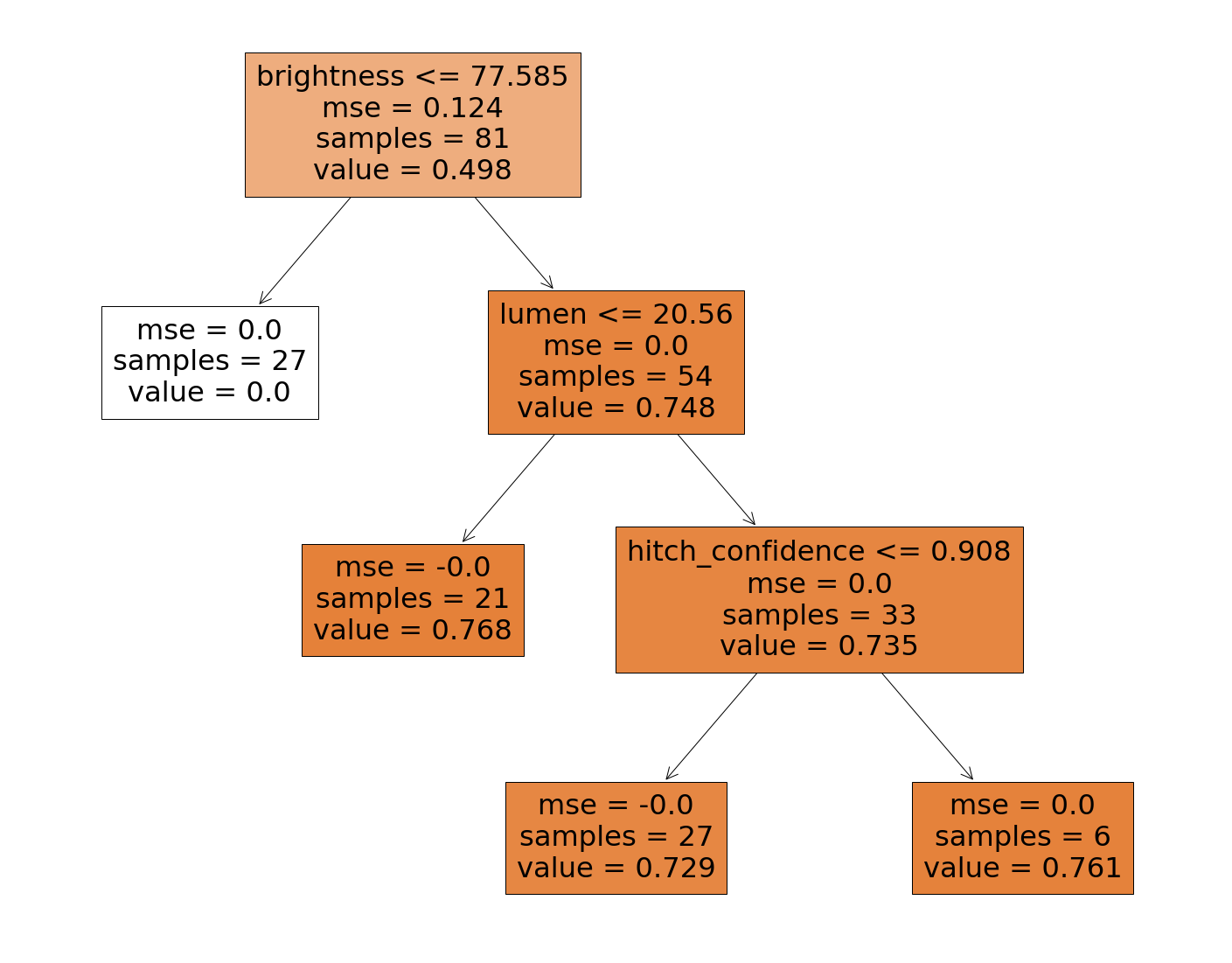}
     \caption{decision tree confidence}
     \label{fig_example_selfx}
  \end{figure} 
  \begin{figure}[!ht]
     \centering
     \includegraphics[width=0.95\columnwidth]{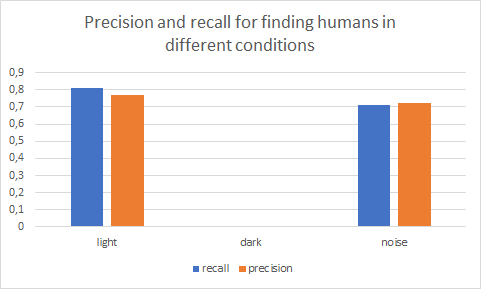}
     \caption{Precision and recall visial search.}
     \label{fig_example_selfx}
  \end{figure} 

\section{Conclusions}
\label{Section_Conclusions} 

There are various bibliography styles available. You can select the style of your choice in the preamble of this document. These styles are Elsevier styles based on standard styles like Harvard and Vancouver. Please use Bib\TeX\ to generate your bibliography and include DOIs whenever available.

\section*{Acknowledgement}
This work was conducted under the RVO program Knowledge Management in Autonomous Systems in cooperation with the ERP Appl.AI.

\bibliography{bibfile_selfx}

\begin{thebibliography}{10}
\expandafter\ifx\csname url\endcsname\relax
  \def\url#1{\texttt{#1}}\fi
\expandafter\ifx\csname urlprefix\endcsname\relax\def\urlprefix{URL }\fi
\expandafter\ifx\csname href\endcsname\relax
  \def\href#1#2{#2} \def\path#1{#1}\fi

\bibitem{huizing}
A.~Huizing, C.~Veenman, M.~Neerincx, J.~Dijk, Hybrid ai: The way forward in ai
  by developing four dimensions, in: F.~Heintz, M.~Milano,
  B.~O{\textquoteright}Sullivan (Eds.), Trustworthy AI – Integrating
  Learning, Optimization and Reasoning, Lecture Notes in Computer Science
  (including subseries Lecture Notes in Artificial Intelligence and Lecture
  Notes in Bioinformatics), Springer, 2021, pp. 71--76, 1st International
  Workshop on Trustworthy AI – Integrating Learning, Optimization and
  Reasoning, TAILOR 2020 held as a part of European Conference on Artificial
  Intelligence, ECAI 2020 ; Conference date: 04-09-2020 Through 05-09-2020.
\newblock \href {http://dx.doi.org/10.1007/978-3-030-73959-1_6}
  {\path{doi:10.1007/978-3-030-73959-1_6}}.

\bibitem{Russel}
S.~Russel, Human compatibale: AI and the problem of control, 1st Edition,
  Viking, 2019.

\bibitem{Pickery}
A.~Pickery, The cybernetic brain: sketches of another future, The name of the
  publisher, 2010.

\bibitem{Heylighen}
F.~Heylighen, From Human Computation to the Global Brain: The Self-Organization
  of Distributed Intelligence, In: Michelucci P. (eds) Handbook of Human
  Computation. Springer, New York, NY, 2013.

\bibitem{chapman2018engineering}
W.~L. Chapman, A.~T. Bahill, A.~W. Wymore, Engineering modeling and design, CRC
  Press, 2018.

\bibitem{estefan2007survey}
J.~A. Estefan, et~al., Survey of model-based systems engineering (mbse)
  methodologies, Incose MBSE Focus Group 25~(8) (2007) 1--12.

\bibitem{dori2016model}
D.~Dori, et~al., Model-based systems engineering with OPM and SysML, Vol.~15,
  Springer, 2016.

\bibitem{friedenthal2014practical}
S.~Friedenthal, A.~Moore, R.~Steiner, A practical guide to SysML: the systems
  modeling language, Morgan Kaufmann, 2014.

\bibitem{Bruyninckx_MBSE}
N.~Hochgeschwender, L.~Gherardi, A.~Shakhirmardanov, G.~K. Kraetzschmar,
  D.~Brugali, H.~Bruyninckx, A model-based approach to software deployment in
  robotics, in: 2013 IEEE/RSJ International Conference on Intelligent Robots
  and Systems, 2013, pp. 3907--3914.
\newblock \href {http://dx.doi.org/10.1109/IROS.2013.6696915}
  {\path{doi:10.1109/IROS.2013.6696915}}.

\bibitem{Lind94}
L.~Morten, Modeling goals and functions of complex industrial plants, Applied
  Artificial Intelligence 8~(2) (1994) 259--283.
\newblock \href {http://dx.doi.org/10.1080/08839519408945442}
  {\path{doi:10.1080/08839519408945442}}.

\bibitem{Carlos_Tomasys_2018}
C.~Hernández~Corbato, J.~Bermejo, R.~Sanz, A self-adaptation framework based
  on functional knowledge for augmented autonomy in robots, Integrated
  Computer-Aided Engineering 25 (2018) 1--16.
\newblock \href {http://dx.doi.org/10.3233/ICA-180565}
  {\path{doi:10.3233/ICA-180565}}.

\bibitem{WeynsArxiv}
D.~Weyns, N.~Bencomo, R.~Calinescu, J.~C{\'{a}}mara, C.~Ghezzi, V.~Grassi,
  L.~Grunske, P.~Inverardi, J.~J{\'{e}}z{\'{e}}quel, S.~Malek, R.~Mirandola,
  M.~Mori, G.~Tamburrelli, \href{http://arxiv.org/abs/1903.04771}{Perpetual
  assurances for self-adaptive systems}, CoRR abs/1903.04771.
\newblock \href {http://arxiv.org/abs/1903.04771} {\path{arXiv:1903.04771}}.
\newline\urlprefix\url{http://arxiv.org/abs/1903.04771}

\bibitem{LemosGiese}
R.~Lemos, H.~Giese, H.~Müller, J.~Andersson, M.~Litoiu, B.~Schmerl, G.~Tamura,
  N.~Villegas, T.~Vogel, D.~Weyns, L.~Baresi, B.~Becker, N.~Bencomo, Y.~Brun,
  B.~Cukic, R.~Desmarais, S.~Dustdar, G.~Engels, J.~Wuttke, Software
  Engineering for Self-Adaptive Systems: A Second Research Roadmap, 2013, pp.
  1--32.
\newblock \href {http://dx.doi.org/10.1007/978-3-642-35813-5_1}
  {\path{doi:10.1007/978-3-642-35813-5_1}}.

\bibitem{IftikharWeyns}
I.~U, D.~Weyns, Activforms: Active formal models for self-adaptation, in: In
  Proc. of the Int. Conf. on Software Engineering for Adaptive and
  Self-Managing Systems (SEAMS2014), 2014, pp. 125 -- 134.

\bibitem{ZhangCheng}
J.~Zhang, B.~H.~C. Cheng, Model-based development of dynamically adaptive
  software, in: In Proc. of the Int. Conf. on Software Engineering, 2006, pp.
  371 -- 380.

\bibitem{MAPE_K}
J.~O. {Kephart}, D.~M. {Chess}, The vision of autonomic computing, Computer
  36~(1) (2003) 41--50.
\newblock \href {http://dx.doi.org/10.1109/MC.2003.1160055}
  {\path{doi:10.1109/MC.2003.1160055}}.

\bibitem{Schlenoff_4DRCS}
C.~Schlenoff, J.~Albus, E.~Messina, A.~Barbera, R.~Madhavan, S.~Balakirsky,
  Using 4d/rcs to address ai knowledge integration, AI Mag. 27 (2006) 71--81.

\bibitem{CORA_2013}
E.~Prestes, J.~Carbonera, S.~Fiorini, V.~Jorge, M.~Abel, R.~Madhavan,
  A.~Locoro, P.~Gonçalves, M.~Barreto, M.~Habib, A.~Chibani, S.~Gérard,
  Y.~Amirat, C.~Schlenoff, Towards a core ontology for robotics and automation,
  Robotics and Autonomous Systems 61 (2013) 1193--1204.
\newblock \href {http://dx.doi.org/10.1016/j.robot.2013.04.005}
  {\path{doi:10.1016/j.robot.2013.04.005}}.

\bibitem{Neto_CORA_2019}
A.~B. d.~O.~{Neto}, J.~A. {Silva}, M.~E. {Barreto}, Prototyping and validating
  the cora ontology: Case study on a simulated reconnaissance mission, in: 2019
  Latin American Robotics Symposium (LARS), 2019 Brazilian Symposium on
  Robotics (SBR) and 2019 Workshop on Robotics in Education (WRE), 2019, pp.
  341--345.

\bibitem{CORA_TO_2017}
S.~Balakirsky, C.~Schlenoff, S.~R. Fiorini, S.~Redfield, M.~Barreto,
  H.~Nakawala, J.~L. Carbonera, L.~Soldatova, J.~Bermejo-Alonso, F.~Maikore,
  P.~J. Goncalves, E.~De~Momi, V.~R. Kumar, T.~Haidegger, Towards a robot task
  ontology standard, in: Proceedings of the ASME 2017 12th International
  Manufacturing Science and Engineering Conference collocated with the
  JSME/ASME 2017 6th International Conference on Materials and Processing.,
  Vol.~3, 2019.
\newblock \href {http://dx.doi.org/10.1115/MSEC2017-2783}
  {\path{doi:10.1115/MSEC2017-2783}}.

\bibitem{Waterloo_Part1}
K.~Czarnecki, Operational world model ontology for automated driving systems
  – part 1: Road structure, Tech. rep., Waterloo Intelligent Systems
  Engineering Lab (2018).
\newblock \href {http://dx.doi.org/10.13140/RG.2.2.15521.30568}
  {\path{doi:10.13140/RG.2.2.15521.30568}}.

\bibitem{Waterloo_Part2}
K.~Czarnecki, Operational world model ontology for automated driving systems
  – part 2: Road users, animals, other obstacles, and environmental
  conditions, Tech. rep., Waterloo Intelligent Systems Engineering Lab (2018).
\newblock \href {http://dx.doi.org/10.13140/RG.2.2.11327.00165}
  {\path{doi:10.13140/RG.2.2.11327.00165}}.

\bibitem{Thai_Grun_2020}
J.~Thai, M.~Gruninger, Robot meets world, in: In Proc. of the Joint Ontology
  Workshop, JOWO., 2020.

\bibitem{KnowRob2010}
M.~Tenorth, D.~Jain, M.~Beetz, Knowledge processing for cognitive robots,
  KI-Kunstliche Intelligenz 24 (2010) 233--340.

\bibitem{KnowRob2018}
M.~Beetz, D.~Beßler, A.~Haidu, M.~Pomarlan, A.~Bozcuoglu, G.~Bartels, Know rob
  2.0—a 2nd generation knowledge processing framework for cognition-enabled
  robotic agents, in: In Proc. of the 2018 IEEE Int. Conf. on Robotics and
  Automation (ICRA), 2018, pp. 512--519.

\bibitem{ScioniBruyninckx2016}
E.~Scioni, N.~H\"{u}bel, S.~Blumenthal, A.~Shakhimardanov, M.~Klotzb\"{u}cher,
  H.~Garcia, H.~Bruyninckx, Hierarchical hypergraphs for knowledge-centric
  robot systems: a composable structural meta model and its domain specific
  language {NPC4} 7~(1) (2016) 55--74.

\bibitem{Lewis_2019}
N.~Taherinejad, P.~Lewis, A.~Jantsch, A.~M. Rahmani, L.~Esterle, Resource
  constrained self-aware cyber-physical systems (tutorial), 2019.
\newblock \href {http://dx.doi.org/10.1109/FAS-W.2019.00071}
  {\path{doi:10.1109/FAS-W.2019.00071}}.

\bibitem{FamSec}
B.~W. Israelsen, N.~R. Ahmed, E.~W. Frew, D.~Lawrence, B.~Argrow,
  \href{http://arxiv.org/abs/1810.06519}{Machine self-confidence in autonomous
  systems via meta-analysis of decision processes}, CoRR abs/1810.06519.
\newblock \href {http://arxiv.org/abs/1810.06519} {\path{arXiv:1810.06519}}.
\newline\urlprefix\url{http://arxiv.org/abs/1810.06519}

\bibitem{Sweet}
N.~Sweet, N.~Ahmed, U.~Kuter, C.~Miller, Towards self-confidence in autonomous
  systems, 2016.
\newblock \href {http://dx.doi.org/10.2514/6.2016-1651}
  {\path{doi:10.2514/6.2016-1651}}.

\bibitem{Carlos_MetaControl2020}
C.~Hernández~Corbato, Z.~Milosevic, C.~Olivares, G.~Rodriguez, C.~Rossi,
  Meta-control and Self-Awareness for the UX-1 Autonomous Underwater Robot,
  2020, pp. 404--415.
\newblock \href {http://dx.doi.org/10.1007/978-3-030-35990-4_33}
  {\path{doi:10.1007/978-3-030-35990-4_33}}.

\bibitem{Lewis_2015}
P.~Lewis, A.~Chandra, F.~Faniyi, K.~Glette, T.~Chen, R.~Bahsoon, J.~Torresen,
  Architectural aspects of self-aware and self-expressive computing systems:
  From psychology to engineering, Computer 48 (2015) 62--70.
\newblock \href {http://dx.doi.org/10.1109/MC.2015.235}
  {\path{doi:10.1109/MC.2015.235}}.

\bibitem{Bermejo_Oasys_2010}
J.~Bermejo, R.~Sanz, M.~Rodriguez, C.~Hernández~Corbato, An ontological
  framework for autonomous systems modelling 3.

\bibitem{Tas_Robustsense_2017}
O.~Tas, S.~Hoermann, B.~Schäufele, F.~Kuhnt, Automated vehicle system
  architecture with performance assessment, 2017.
\newblock \href {http://dx.doi.org/10.1109/ITSC.2017.8317862}
  {\path{doi:10.1109/ITSC.2017.8317862}}.

\bibitem{Waymore}
W.~A. Waymore, Model-based systems engineering: An introduction to the
  mathematical theory of discrete systems and to the tricotyledon theory of
  system design, 1st Edition, CRC Press, 1993.

\bibitem{bjelonicYolo2018}
M.~Bjelonic, {YOLO ROS}: Real-time object detection for {ROS},
  \url{https://github.com/leggedrobotics/darknet_ros} (2016--2018).

\bibitem{spot}
B.~Dynamics, \href{https://www.bostondynamics.com/spot}{Spot}.
\newline\urlprefix\url{https://www.bostondynamics.com/spot}

\bibitem{respeaker}
S.~Studio, \href{https://wiki.seeedstudio.com/ReSpeaker_Core_v2.0/}{Respeaker
  core v2.0}.
\newline\urlprefix\url{https://wiki.seeedstudio.com/ReSpeaker_Core_v2.0/}

\bibitem{Grakn_blog}
H.~Pribadi,
  \href{https://blog.grakn.ai/the-grakn-ai-ontology-simplicity-and-maintainability-ab78340f5ff6}{The
  grakn.ai ontology: simplicity and maintainability, in comparison with
  traditional ontology languages and tools} (2021).
\newline\urlprefix\url{https://blog.grakn.ai/the-grakn-ai-ontology-simplicity-and-maintainability-ab78340f5ff6}

\bibitem{Sijs_iros}
J.~Sijs, J.~Fletcher, An online knowledge base to model real-world, indoor
  environments for robots conducting a search operation with uncertainty, in:
  submitted to the 2021 IEEE/RSJ International Conference on Intelligent Robots
  and Systems (IROS'21)., 2021.

\bibitem{brisque}
A.~Mittal, A.~K. Moorthy, A.~C. Bovik, Blind/referenceless image spatial
  quality evaluator, in: 2011 Conference Record of the Forty Fifth Asilomar
  Conference on Signals, Systems and Computers (ASILOMAR), 2011, pp. 723--727.
\newblock \href {http://dx.doi.org/10.1109/ACSSC.2011.6190099}
  {\path{doi:10.1109/ACSSC.2011.6190099}}.

\end{thebibliography}

\section*{Appendices}
  \begin{figure*}[t]
     \centering
     \begin{subfigure}[b]{0.8\textwidth}
         \centering
         \includegraphics[width=1.0\textwidth]{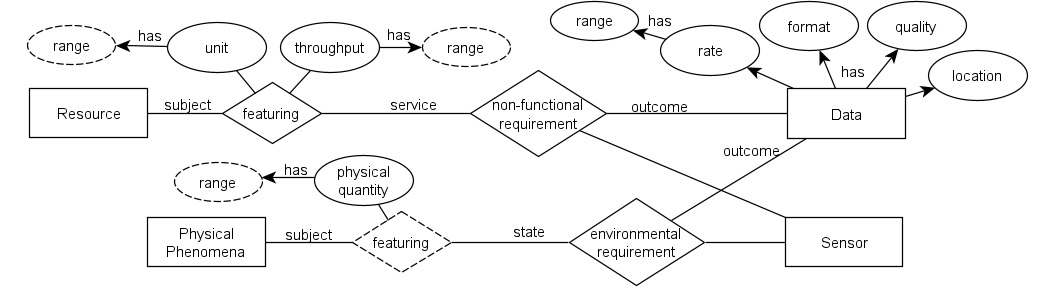}
         \caption{Model for a  sensr unit}
         \label{fig_sensor_unit}
     \end{subfigure}
     \begin{subfigure}[b]{0.8\textwidth}
         \centering
         \includegraphics[width=1.0\textwidth]{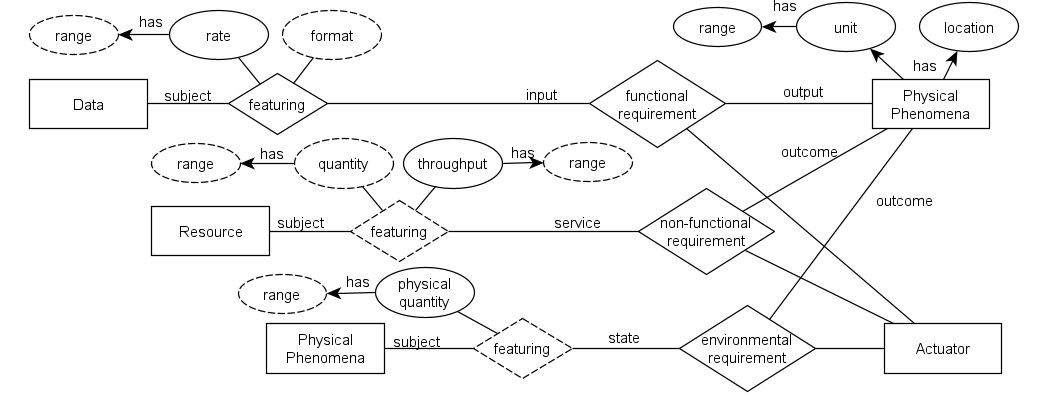}
         \caption{Model of an actuator unit}
         \label{fig_actuator_unit}
     \end{subfigure}
     \begin{subfigure}[b]{0.8\textwidth}
         \centering
         \includegraphics[width=1.0\textwidth]{figures/functional_unit.jpg}
         \caption{Model of a functional unit}
         \label{fig_functional_unit}
     \end{subfigure}
     \begin{subfigure}[b]{0.8\textwidth}
         \centering
         \includegraphics[width=1.0\textwidth]{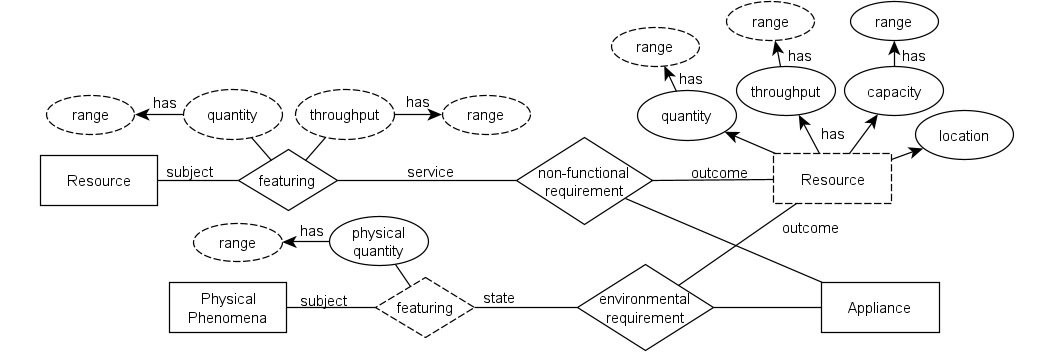}
         \caption{Model of an appliance unit}
         \label{fig_appliance_unit}
     \end{subfigure}
        \caption{The four different types of Units that are modelled for describing a system configuration.}
        \label{fig_unit_types}
  \end{figure*}   

\end{document}